\documentclass[12pt]{article}
\usepackage[margin=2 cm]{geometry}
\usepackage[utf8]{inputenc}
\usepackage[nottoc]{tocbibind}
\usepackage{comment}
\usepackage{graphicx}
\usepackage{amsmath}
\usepackage{hyperref}
\usepackage{amsmath,amssymb,extarrows,mathtools,graphicx,subfigure,setspace}
\usepackage{cite}
\usepackage{amsmath}
\usepackage{braket}
\usepackage{epsfig}
\usepackage[section]{placeins}
\usepackage{slashed}
\usepackage{color}
\usepackage{caption}
\usepackage[font=small,labelfont=bf]{caption}
\makeatother

\newcommand{\be}{\begin{equation}}
\newcommand{\bea}{\begin{eqnarray}}
\newcommand{\eea}{\end{eqnarray}}
\newcommand{\ba}{\begin{array}}
\newcommand{\ea}{\end{array}}
\newcommand{\ee}{\end{equation}}
\newcommand{\bes}{\begin{equation*}}
\newcommand{\beas}{\begin{eqnarray*}}
\newcommand{\eeas}{\end{eqnarray*}}
\newcommand{\bas}{\begin{array*}}
\newcommand{\eas}{\end{array*}}
\newcommand{\ees}{\end{equation*}}

\numberwithin{equation}{section}

\begin{document}
\onehalfspacing
\vfill
\begin{titlepage}
\vspace{10mm}

\begin{center}

\vspace*{10mm}
\vspace*{1mm}
{\Large  \textbf{Schwinger Effect and Entanglement Entropy\\ in Confining Geometries }} 
 \vspace*{1cm}
 
{ Mahdis Ghodrati}

\vspace*{8mm}
{\textsl{University of Michigan, Randall laboratory of Physics, Ann Arbor, MI 48109-1040, USA}}

\textsl{E-mail: {ghodrati@umich.edu}}
 \vspace*{2mm}

\vspace*{2cm}

\end{center}

\begin{abstract}
By using AdS/CFT, we study the critical electric field, the Schwinger pair creation rate and the potential phase diagram for the quark and anti quark in four confining supergravity backgrounds which are the Witten QCD, the Maldacena-Nunez, the Klebanov-Tseytlin and the Klebanov-Strassler models. We compare the rate of phase transition in these models and compare it also with the conformal case.
We then present the phase diagrams of the entanglement entropy of a strip in these geometries and find the predicted butterfly shape in the diagrams. We find that the phase transitions have higher rate in WQCD and KT relative to MN and KS. Finally we show the effect of turning on an additional magnetic field on the rate of pair creation by using the imaginary part of the Euler-Heisenberg effective Lagrangian. The results is increasing the parallel magnetic field would increase the pair creation rate and increasing the perpendicular magnetic field would decrease the rate.  
\end{abstract}

\end{titlepage}

\tableofcontents

%%%%%%%%%%%%%%%%%%%
%%%%%%%%%%%%%%%%%%%%%%%%%%%%%%%%%%%%%%%%%%%%%%%%%%%%%%%%%%%%%%%%%%%%%%%%%%%%%%
%%%%%%%%%%%%%%%%%%%%%%%%%%%%%%%%%%%%%%%%%%%%%%%%%%%%%%%%%%%%%%%%%%%%%%%%%%%%%%
\section{Introduction}

Schwinger effect in quantum field theory \cite{Schwinger:1951nm} is the creation of pairs of particles in the presence of a strong electric or magnetic field. Intuitively when the strength of the electric field reaches the string tension between the quark anti-quark, it can break the string that is attaching them and so the virtual pairs can become on-shell and a current can be created. 
Semenoff and Zarembo \cite{Semenoff:2011ng} first studied the Schwinger effect holographically. Then the authors in \cite{Hubeny:2014kma} and \cite{Hubeny:2014zna}, using the Nambu-Goto action on a probe brane and by calculating the free energy and using the first law of thermodynamic, calculated the entanglement entropy of a quark and an anti-quark which are accelerating in an electric field in the $\text{AdS}_5$ background. A similar calculation in a different setup was also done in \cite{Lewkowycz:2013laa}. So one might want to try this in other backgrounds and specifically in the confining ones. 

On the other hand, there are some ideas on how to measure the entanglement entropy in the lab by using the fluctuations of current which is flowing through a quantum point contact as the probe, \cite{Klich:2008un} \cite{Klich:2011eg}, which still are not quite successful.  Another conjecture is that there might be a relation between the Schwinger effect and the entanglement entropy. Finding such a relation could be a breakthrough since the Schwinger pair creation rate could also act as an entanglement meter in the condensed matter systems. To explore for such a relation, in the first step, we study the phase diagrams of Schwinger effect and of entanglement entropy for four confining geometries; the Witten-QCD, the Maldacena-Nunez, the Klebanov-Strassler and the Klebanov-Tseytlin which are dual to $\mathcal{N}=1$ field theories. 

For calculating the entanglement entropy for accelerating particles one can use the method in\cite{Hubeny:2014zna}. For doing so, first one needs to find the string world sheet profile in these confining backgrounds. For the case of $\text{AdS}$ geometry, due to the large amount of symmetries, the corresponding PDE equation of motion is simple and has been solved in \cite{Xiao:2008nr}.  Mikhailov also found a simple linear relation \cite{Mikhailov:2003er} to find the string world sheet profile based on the position of the quark and anti-quark on the boundary that can only be used for the $\text{AdS}_5$ background. For other geometries such as the confining backgrounds there is no Mikhailov-like equation and for finding the string profile one needs to solve several more difficult PDE equations analytically which for our supergravity background geometries we present them in section 2.  We just do the similar calculation for the Minkowski background and we find the free energy of the quarks and anti quarks in a flat background in section 3.

In order to look for a relationship between the phase transitions of Schwinger effect and the entanglement entropy, we study the phase transitions in an electric potential similar to the procedures in \cite{Sato:2013dwa}\cite {Sato:2013hyw}\cite{Sato:2013iua} \cite{Sato:2013pxa}\cite{Kawai:2015mha}\cite{Kawai:2013xya} in section 4. By using DBI action, we calculate the critical electric fields in these geometries where the potential becomes catastrophically unstable and we then find the phase diagrams numerically. Interestingly the phase diagrams of all of these geometries are very similar where three different phases can be detected as has been predicted in \cite{Sato:2013hyw} as a universal feature of all confining geometries. We also compare the diagrams with the conformal case of Klebanov-Witten which only two phases can be detected.

In section 5 we look at the phase diagrams of the entanglement entropy of a strip, similar to the calculation in  \cite{Klebanov:2007ws} , \cite{Kol:2014nqa} and \cite{Faraggi:2007fu}. Klebanov, Kutasov and Murugan found a generalization of Ryu-Takayanagi relation for the non-conformal geometries \cite{Klebanov:2007ws}. Then in \cite{Kol:2014nqa}, the authors presented the plots of the phase diagram for geometries constructed by Dp brane compactified on a circle which are the generalization of the Witten-QCD model. They have found a butterfly shape and a double valuedness in the phase diagram of these confining geometries. In this paper, we additionally present the phase diagrams of the Maldacena-Nunez, Klebanov-Strassler and Klebanov-Tseytlin plus the Klebanov-Witten and Witten-QCD models by similarly calculating the length of the connected regions and the entanglement entropy of the connected and disconnect solutions. We also find the similar butterfly shape in the phase diagrams. We also find that if in a specific geometry the phase transition of Schwinger effect is fast and dramatic, this also would be the case in its entanglement entropy phase transition which is the case for WQCD and KT, and if the phase transition for Schwinger effect be mild, this also would be the case for its EE phase diagram which is the case for MN and KS. One can also compare the features with the conformal case of Klebanov-Witten and AdS as a limit of a mild transition. 

In section 6, we study the Schwinger effect in the presence of a magnetic field in addition to the electric field and study its effects on the pair creation rate. By adding a probe D8-brane, in \cite{Hashimoto:2013mua} and \cite{Hashimoto:2014yya} the authors studied the imaginary part of the Euler-Heisenberg effective Lagrangian, the rate of pair creation, the critical electric field and the effect of the parallel and perpendicular components of the magnetic field on the rate of pair creation in the background of Sakai-Sugimoto and the deformed Sakai-Sugimoto models. Similarly we calculate the DBI action and the imaginary part of the Euler-Heisenberg effective Lagrangian in our geometries. We find that in all of our supergravity confining geometries the parallel magnetic field would increase the rate of pair creation while the perpendicular magnetic field would decrease it which should be a general feature of all confining backgrounds. We then conclude with a discussion in section 7.

\section{The string profile in confining geometries }
In this section we present the PDE equations of the string profiles which can be used in finding the entanglement entropy of accelerating quarks moving on specific trajectories.

As in \cite{Xiao:2008nr}, by starting from the Nambu-Goto action for the $\text{AdS}_5$ geometry,
\begin{gather}
ds^2=R^2 [\frac{du^2}{u^2}-u^2 dt^2+u^2(dx^2+dy^2+dz^2)],
\end{gather}
and then by assuming the static gauge of $(\tau, \sigma)=(t,u)$ and the embedding coordinate of $X^\mu=(t,u,x(t,u),0,0)$, one can find the determinant of the induced metric as
\begin{gather}
\sqrt{-g} =R^2 \sqrt{1-\dot{x}^2+u^4 x'^2},
\end{gather}
and the equation of motion as
\begin{gather}
\frac{\partial}{\partial u} \left(\frac{u^4 x'}{\sqrt{-g}}\right)-\frac{\partial}{\partial t}\left(\frac{\dot{x}}{\sqrt{-g}}\right)=0.
\end{gather}
Solving these PDE equations, in general is a difficult task, but for the case of $\text{AdS}$ which enjoys a lot of symmetries and has a simpler form, the author in \cite{Xiao:2008nr} could find the solution and therefore the string profile as a function of $t$ and $u$ as
\begin{gather}
x=\pm \sqrt{t^2+b^2-\frac{1}{u^2}}.
\end{gather}
If a heavy quark and an anti-quark accelerate on a specific trajectory due to the potential of an electric field, the classical solution from the Nambu-Goto action is a world sheet that is a part of $\text{AdS}_2$ and the locus of
\begin{gather}
u^2+(x^1)^2-(x^0)^2=\frac{M^2}{E^2}.
\end{gather}
where $M$ is the mass of the quark and $E$ is the electric field. From this relation, the world sheet event horizon can be read as $u_E=\frac{M}{E}$. As in \cite{Hubeny:2014zna}, for a specific trajectory such as a hyperbola, one can read the metric near the quark trajectory.

Alternatively for finding the induced metric on the world sheet, similar to \cite{Hubeny:2014kma}, one can use the Mikhailov relation between the embedding coordinate $X^M(\tau,u)$ and the boundary quark position $x^\mu(\tau)$, \cite{Mikhailov:2003er}, as
\begin{gather}
X^\mu(\tau,u)=u { \dot{x}}^\mu (\tau)+{x^\mu} (\tau), \ \ \ \ \ \ X^M(\tau,u)=(X^\mu(\tau,u),u),
\end{gather}

Then as in \cite{Hubeny:2014zna} one can find the proper area between the probe brane and the event horizon which is proportional to the free energy. By knowing the Unruh temperature that the quarks would feel in the accelerated reference frame in $\text{AdS}$ as $T_U=\frac{E}{2\pi M}$, and by using the first law of thermodynamics one can read the entropy. If one can assume the semi-classical and heavy quark limit in the problem, all of those entanglement would be due to the ``entanglement entropy" and so  for the AdS the EE where found to be $s=\sqrt{\lambda}$,  \cite{Hubeny:2014zna} where $\lambda$ is the 't Hooft coupling constant.  

Now we look at the equations for the $\mathcal{N}=1$ supergravity solutions that are dual to the confining geometries. One should notice that for this calculation, one first needs the string world sheet profile, the induced metric and the Unruh temperature of the accelerating particles in each background which we don't present here.

\subsection{Witten-QCD}
Between all the three mentioned confining geometries, the Witten QCD model is the most similar background to the AdS metric. In the string frame, its metric and dilaton field are \cite{Witten:1997ep},
\begin{gather}
ds^2=(\frac{u}{R})^{3/2}\left(\eta_{\mu\nu} dx^\mu dx^\nu+\frac{4R^3}{9u_0} f(u) d\theta^2\right)+\left(\frac{R}{u}\right)^{3/2} \frac{du^2}{f(u)}+R^{3/2} u^{1/2} d\Omega_4^2,\nonumber\\
f(u)=1-\frac{u_0^3}{u^3}, \ \ \ \ \ \ \ \ R=(\pi Ng_s)^{\frac{1}{3}} {\alpha^\prime}^{\frac{1}{2}},\ \ \ \ \ \
e^\Phi=g_s \frac{u^{3/4}}{R^{3/4}}.
\end{gather}
If we consider the static gauge, and the embedding coordinate as $X^\mu=(t,u,x(t,u),0,0,\theta(t,u),0,0,0,0)$, then the equations would be complicated. So we assume $\theta$ is a constant, and therefore the determinant of the induced metric simplifies to
\begin{gather}
\sqrt{-g}=\sqrt{u^3 \left(\frac{ 1-\dot{x}^2}{u^3-{u_0}^3}+\frac{{x'}^2}{R^3 }\right)},
\end{gather}
which leads to the following equation of motion,
\begin{gather}
\left(\frac{u^3-u_0^3}{R^3 u^3} \right)\frac{\partial}{\partial u}\left(\frac{u^3 x'}{\sqrt{-g}}\right)-\frac{\partial }{\partial t}\left(\frac{\dot{x}}{ \sqrt{-g}}\right)=0,
\end{gather}
which is quit similar to the AdS case. If one finds the analytical solution of this PDE equation, similar to the AdS case, one can follow the procedures of \cite{Hubeny:2014zna} and find the entanglement entropy of heavy accelerating quarks in this model. 

\subsection{Maldacena-Nunez}
The MN metric is obtained by a large number of D5 branes wrapping on $S^2$, \cite{Maldacena:2000yy}. In the string frame the metric and the fields are \cite{Basu:2012ae}
\begin{gather}
ds^2_{10}=e^\phi [ -dt^2+{dx_1}^2+{dx_2}^2+{dx_3}^2+e^{2h(r)}({d\theta_1}^2+\sin^2 \theta_1 {d\phi_1}^2)+dr^2+\frac{1}{4}(w^i-A^i)^2 ],
\end{gather}
where
\begin{gather}
A^1=-a(r)d\theta_1, \ \ \ \ \ \ \ \ A^2=a(r)\sin \theta_1 d\phi_1, \ \ \ \ \ \ \ \ 
A^3= -\cos \theta_1d\phi_1,
\end{gather}
and the $\omega^i$ 's parameterize the compactification 3-sphere which are
\begin{gather}
\omega^1 =\cos \psi d\theta_2+\sin \psi \sin\theta_2 d \phi_2,\ \ \ \ \ \ \ \ \ 
\omega^2 =-\sin \psi d \theta_2+\cos \psi \sin \theta_2 d\phi_2\nonumber\\
\omega^3=d\psi+\cos \theta_2 d\phi_2,
\end{gather}
and also the other parameters of the metric are
\begin{gather}
a(r)=\frac{2r}{\sinh 2r}, \ \ \ \ 
e^{2h}= r\coth {2r}-\frac{r^2}{{\sinh 2r}^2}-\frac{1}{4},\ \ \ \ 
e^{-2\phi} =e^{-2\phi_0} \frac{2 e^h}{\sinh 2r}.
\end{gather}

For the case of Maldacena-Nunez model, we assume the embedding coordinate as $(t,r, x(t,r))$ and all other coordinates will set to be zero. Then one would get
\begin{gather}
\sqrt{-g}=e^\phi \sqrt{(1-\dot{x}^2 )(1+x'^2)}.
\end{gather}
So the equation of motion of the string profile in this background is
\begin{gather}
\frac{\partial}{\partial r} \left( \frac{e^{2\phi} x'  (1-\dot{x}^2 )}{\sqrt{-g } }\right)-\frac{\partial}{\partial t} \left( \frac{ e^{2\phi} \dot{x} (1+x'^2) }{\sqrt{ (x'^2+1)(\dot{x}^2-1) } }  \right)=0.
\end{gather}

Again by solving this equation analytically one can find the string profile and then the entanglement entropy of heavy accelerating quarks.

\subsection{Klebanov-Tseytlin}
The Klebanov-Tseytin metric is a singular solution which is dual to the chirally symmetric phase of the Klebanov-Strassler model which has D3 brane charges that dissolve in the flux \cite{Bena:2012ek}.

The metric is
\begin{gather}
ds_{10}^2=h(r)^{-1/2} \left[ -dt^2+d \vec{x}^2 \right]+h(r)^{1/2} \left[ dr^2+r^2 ds_{T^{1,1} }^2 \right].
\end{gather}
Here $ds_{T^{1,1} }^2$ is a base of a cone with the definition of
\begin{gather}
ds_{T^{1,1} }^2=\frac{1}{9} (g^5)^2+\frac{1}{6} \sum_{i=1}^4 (g^i)^2.
\end{gather}
It is the metric on the coset space $T^{1,1}=(SU(2) \times SU(2))/U(1)$.
Also $g^i$ are some functions of the angles $\theta_1, \theta_2, \phi_1, \phi_2, \psi$ as
\begin{gather}
g^1=(-\sin \theta_1 d\phi_1-\cos \psi \sin \theta_2 d\phi_2 +\sin \psi d\theta_2)/ \sqrt{2}, \ \ \ \ 
g^2=(d\theta_1-\sin \psi \sin \theta_2 d\phi_2-\cos \psi d\theta_2) / \sqrt{2},\nonumber\\
g^3=(-\sin \theta_1 d\phi_1+\cos \psi \sin \theta_2 d \phi_2-\sin\psi d \theta_2) \sqrt{2}, \ \ \ \ \ 
g^4=(d\theta_1+ \sin \psi \sin \theta_2 d\phi_2+\cos \psi d\theta_2) /\sqrt{2},\nonumber\\
g^5=d\psi+\cos \theta_1 d\phi_1+\cos \theta_2 d\phi_2,
\end{gather}
and also
\begin{gather}
h(r)=\frac{L^4}{r^4} \ln{\frac{r}{r_s}}, \ \ \ \ \ \ \ L^4=\frac{81}{2} g_s M^2 \epsilon^4.
\end{gather}
In this frame, the asymptotic flat region has been eliminated. Also $r=r_s$ is where the naked singularity is located.

For the case of KT the embedding would be $(t, \tau, x(t,\tau))$ with all other coordinates zero. Then 
\begin{gather}
\sqrt{-g}=\sqrt{(1-\dot{x}^2)(\frac{{x'}^2}{H}+\frac {\epsilon^2}{9} e^ {\frac{2\tau}{3}}) },
\end{gather} 
and the equation of motion of the string profile would be
\begin{gather}
\frac{\partial}{\partial \tau} \left(\frac{x' (1-\dot{x}^2 )}{H\sqrt{-g}} \right) -\frac{\partial}{\partial t} \left(\frac{\dot{x}  (\frac{x'^2}{H}+\frac{\epsilon^2}{9} e^{\frac{2\tau}{3} } ) }{\sqrt{-g}}  \right)=0.
\end{gather}

\subsection{Klebanov-Strassler}
The Klebanov-Strassler (KS) metric which is known also as warped deformed conifold is obtained by a collection of $N$ regular and $M$ fractional D3-branes  \cite{Klebanov:2000hb}. 

The metric is
\begin{gather}
ds_{10}^2=h^{-\frac{1}{2}}(\tau) dx_\mu dx^\mu+h^{\frac{1}{2}}(\tau) {ds_6}^2,
\end{gather}
and $ds_6^2$ is the metric of the deformed conifold which is
\begin{gather}
{ds_6}^2=\frac{1}{2} \epsilon^{\frac{4}{3}} K(\tau) \Big[ \frac{1}{3 K^3(\tau)} (d\tau^2+ (g^5)^2 )+\cosh^2 (\frac{\tau}{2}) [(g^3)^2+(g^4)^2]+\sinh^2 (\frac{\tau}{2}) [(g^1)^2+(g^2)^2] \Big].
\end{gather}
The parameters of the metric are
\begin{gather}
K(\tau)=\frac{(\sinh(2\tau) -2\tau)^{\frac{1}{3}}} {2^{\frac{1}{3}} \sinh \tau },\ \ \ \ \ \ \ \ \ h(\tau)=(g_s M \alpha')^2 2^{2/3} \epsilon^{-8/3} I(\tau),\nonumber\\  I(\tau)=\int_\tau^\infty dx \frac{x \coth x-1}{\sinh^2 x} (\sinh(2x)-2x)^{\frac{1}{3}}, 
\end{gather}
and
\begin{gather}
g^1=\frac{1}{\sqrt{2}} [ -\sin \theta_1 d\phi_1-\cos \psi \sin\theta_2 d\phi_2+\sin \psi d\theta_2],\ \ \ \ \
g^2=\frac{1}{\sqrt{2}}[d\theta_1 -\sin \psi \sin\theta_2 d\phi_2-\cos\psi d\theta_2],\nonumber\\
g^3=\frac{1}{\sqrt{2}}[-\sin\theta_1 d\phi_1+\cos \psi \sin \theta_2 d\phi_2-\sin\psi d\theta_2],\ \ \ \ \ 
g^4=\frac{1}{\sqrt{2}}[d\theta_1 +\sin \psi\sin\theta_2 d\phi_2+\cos \psi d\theta_2],\nonumber\\
g^5=d\psi+\cos \theta_1 d\phi_1+\cos\theta_2 d\phi_2.
\end{gather}

For the case of Klebanov-Strassler similar to the KT, the embedding would be $(t, \tau, x(t,\tau))$ with all other coordinate zero. Then 
\begin{gather}
\sqrt{-g}=\sqrt{(1-\dot{x}^2) \left(\frac{x'^2}{h} +\frac{\epsilon^{\frac{4}{3}} }{6 K^2(\tau)} \right)},
\end{gather}
and the equation of motion of the string profile would be
\begin{gather}
\frac{\partial}{\partial \tau} \left(\frac{x' (1-\dot{x}^2) }{h\sqrt{-g}} \right) -\frac{\partial}{\partial t} \left(\frac{\dot{x}}{\sqrt{-g}}  (\frac{x'^2}{h} +\frac{\epsilon^{\frac{4}{3}} }{6 K^2(\tau)} ) \right)=0.
\end{gather} 

\subsection{Klebanov-Witten}
The Klebanov-Witten solution is similar to the KT throat solution but with no logarithmic warping \cite{Kaviani:2015rxa}. Unlike the other four mentioned metrics, it is a conformal non-confining geometry. We study this background to compare our results with the conformal case.  

The metric is
\begin{gather}
ds^2=h^{-\frac{1}{2}} g_{\mu\nu} dx^\mu dx^\nu+h^{\frac{1}{2}} (dr^2+r^2 ds_{T^{1,1}} ^2),
\end{gather}
where,
\begin{gather}
h=\frac{L^4}{r^4}, \ \ \ \ \text{and} \ \ \ \ L^4=\frac{27\pi}{4} g_s N (\alpha')^2.
\end{gather}
The embedding would be $(t, r, x(t,r))$ with all other coordinates zero. Then 
\begin{gather}
\sqrt{-g}=\sqrt{(1-\dot{x}^2)(1+\frac{x'^2}{h} )},
\end{gather} 
and the equation of motion of the string profile is
\begin{gather}
\frac{\partial}{\partial r} \left(\frac{x' (1-\dot{x}^2 )}{h\sqrt{-g}} \right) -\frac{\partial}{\partial t} \left(\frac{\dot{x} (1+\frac{x'^2}{h} ) }{\sqrt{-g}}   \right)=0.
\end{gather} 

Therefore again solving this equation would give the string profile and the induced metric near the accelerating particles' trajectory in the background of the conformal KW model.

\section{The free energy of accelerating $q \bar{q} $ in the Miknowski background}

In \cite{Hubeny:2014zna} by minimizing the Nambu-Goto action and by using the solution of the PDE for accelerating particles in $\text{AdS}_5$, the authors found the metric near the quark and anti-quark trajectory as
\begin{gather}
ds_g^2=\sqrt{\lambda} \alpha' \left[ -\left(\frac{1}{u^2}-\frac{E^2}{M^2} \right) d\tau^2-\frac{2}{u^2} d\tau du\right],
\end{gather}
then calculating the proper area between the probe brane and the event horizon yields
\begin{gather}
S_N=-2\left[M-\epsilon M -\frac{M}{2} \right] \tau_P.
\end{gather}
Knowing the Unruh temperature of AdS, $T_U=\frac{E}{2\pi M}$ and the free energy, $\frac{1}{2} \epsilon =M-\frac{M}{2} -\sqrt{\lambda} T_U$, the entanglement entropy of the accelerating quark and anti-quark have found to be $s=\sqrt{\lambda}$.

Now the flat geometry can be the UV limit of the Hard-Wall and Witten-QCD model. Therefore for the flat metric
\begin{gather}
ds^2=\sqrt{\lambda} \alpha' (-dt^2+dx_1^2+dx_2^2+dx_3^2+dx_4^2+du^2),
\end{gather}
one can repeat the calculation of Semenoff and Hubeny. So one would have
\begin{gather}
\gamma_{\tau \tau}=\dot{x}^2-1, \ \ \ \gamma_{uu}=1+x'^2, \ \ \ \ \gamma_{\tau u}=x' \dot{x},
\end{gather}
and then the equation of motion is
\begin{gather}
\partial_u\left(\frac{x'}{\sqrt{ 1-\dot{x}^2+x'^2} }\right)-\partial_t \left(\frac{\dot{x} }{\sqrt{1-\dot{x}^2+x'^2} } \right)=0,
\end{gather}
and the solution of this PDE can be found as $x(t,u)=\sqrt{t^2+b^2-u^2}$, where as in the previous case, for the accelerating quark and anti-quark, the constant is $b=\frac{M}{E}$. From this solution one can see that the world sheet event horizon is at $u_E=\frac{M}{E}$.
Now by using this solution, the components of the induced metric can be found as
\begin{gather}
\gamma_{\tau \tau}=\frac{u^2-b^2}{t^2+b^2-u^2},\ \ \ 
\gamma_{uu}=\frac{t^2+b^2}{t^2+b^2-u^2},\ \ \ 
\gamma_{\tau u}=\frac{t^2+b^2}{t^2+b^2-u^2}.
\end{gather}

Similar to the conditions that have been applied to derive the induced metric of Semenoff' work, one can similarly reach to the following induced metric,
\begin{gather}
ds^2=\sqrt{\lambda} \alpha' \left[\left(1+\frac{u^2 E^2}{M^2} \right)du^2-\frac{2E^2}{M^2}u^2 d\tau du\right].
\end{gather}

The D3-brane is located at $u_M=\frac{\sqrt{\lambda} }{2\pi M}$.
So the proper area between the probe brane and the event horizon is 
\begin{gather}
S_N=2\big[ -\frac{\sqrt{\lambda}}{2\pi} \int_{-\frac{\tau_p}{2}}^{\frac{\tau_p}{2}} d\tau \int_{u_M}^{u_E} \frac{E^2 u^2}{M^2} du+\frac{M}{2}\big] \nonumber\\
=2\tau_p\big [-\frac{\sqrt{\lambda}}{6\pi}\left(\frac{M}{E}-\frac{E^2}{M^2} (\frac{\sqrt{\lambda}}{2\pi M})^3 \right)+\frac{M}{2} \big].
\end{gather}
Assuming that the Unruh temperature of the accelerated frame is $T_U=\frac{E}{2\pi M}$ then the free energy would be,
\begin{gather}
\frac{\epsilon}{2}=\left(-\frac{\sqrt{\lambda}}{12 \pi^2 T}+\frac{T^2 \lambda^2}{12 \pi^2 M^3}+\frac{M}{2}\right).
\end{gather}

So knowing the solution of any of the above PDE equations in any confining geometries can similarly lead to the proper area between the probe brane and the event horizon and yields the free energy of the accelerating $q \tilde{q}$ in those geometries. In a similar way the holographic Schwinger effect were also studied in other geometries such as in de sitter space \cite{Fischler:2014ama}.

\section{Potential analysis of the confining geometries}

One would think that for searching for any possible relationship between the entanglement entropy and the Schwinger pair creation rate, it would be interesting to first find the phase diagrams and the phase transitions for both quantities in few different confining backgrounds.  

\begin{figure}[h!]
\centering
\begin{minipage}{.55\textwidth}
  \centering
  \includegraphics[width=.9\linewidth]{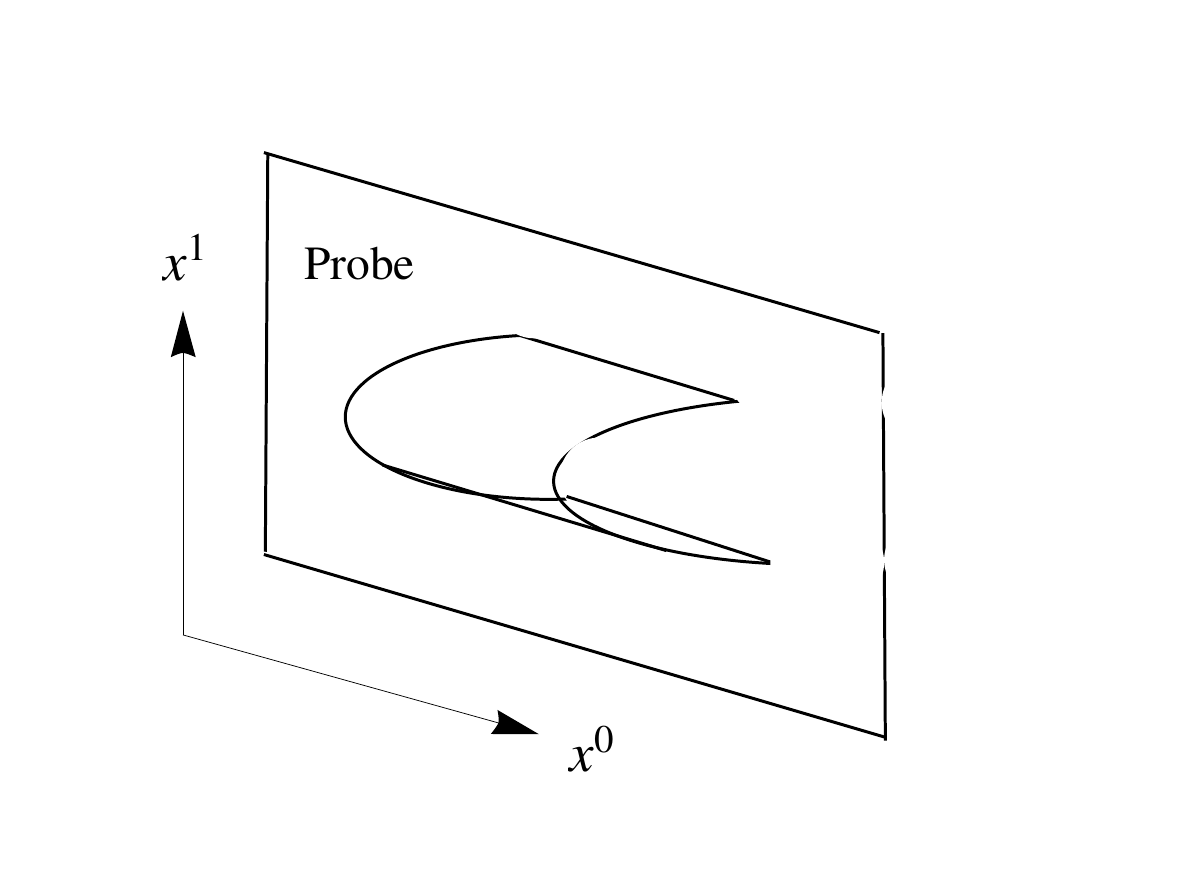}
  \captionof{figure}{World sheet configuration in 3D}
  \label{fig:test13a}
\end{minipage}%
\begin{minipage}{.55\textwidth}
  \centering
  \includegraphics[width=.9\linewidth]{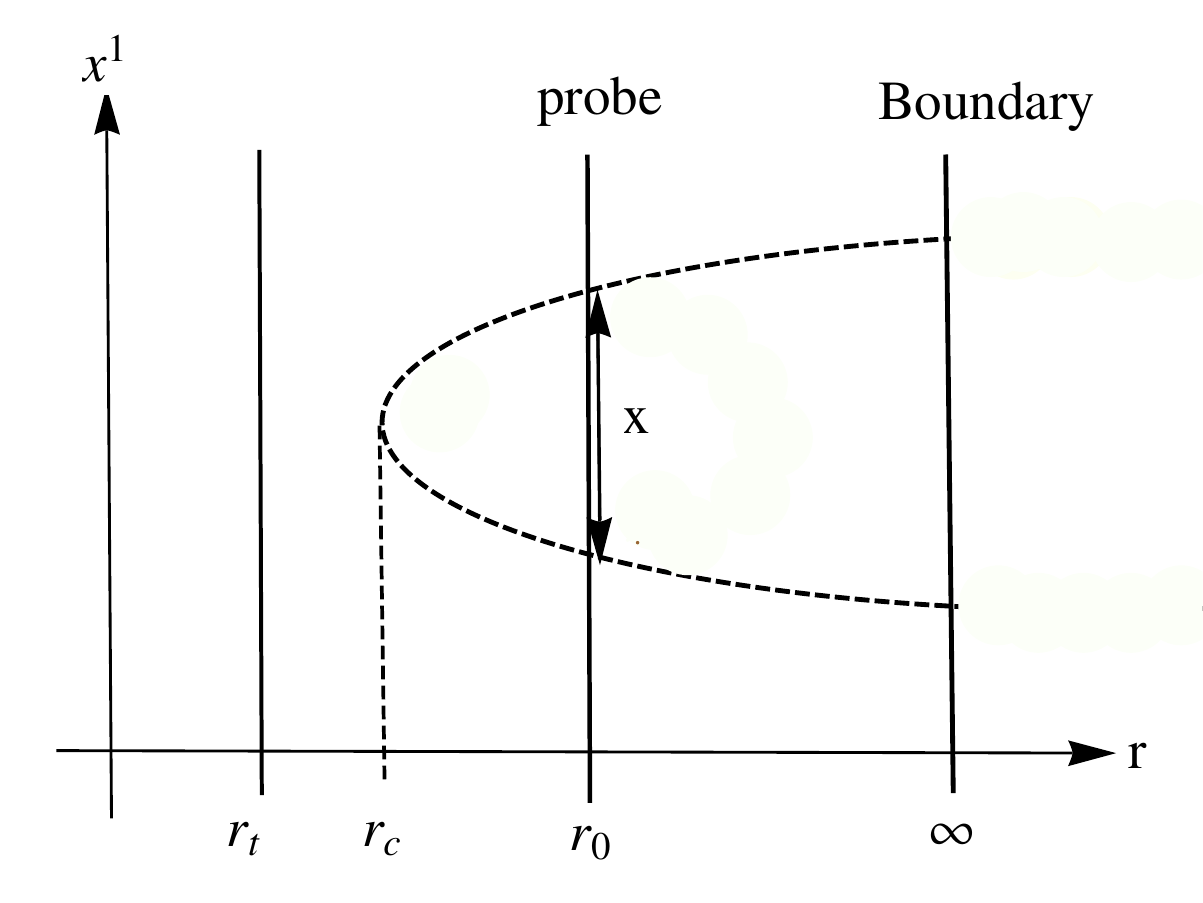}
  \captionof{figure}{World sheet configuration in 2D}
  \label{test14b}
\end{minipage}
\end{figure}

\begin{figure}[h!]
\centering
\begin{minipage}{.5\textwidth}
  \centering
  \includegraphics[width=.9\linewidth]{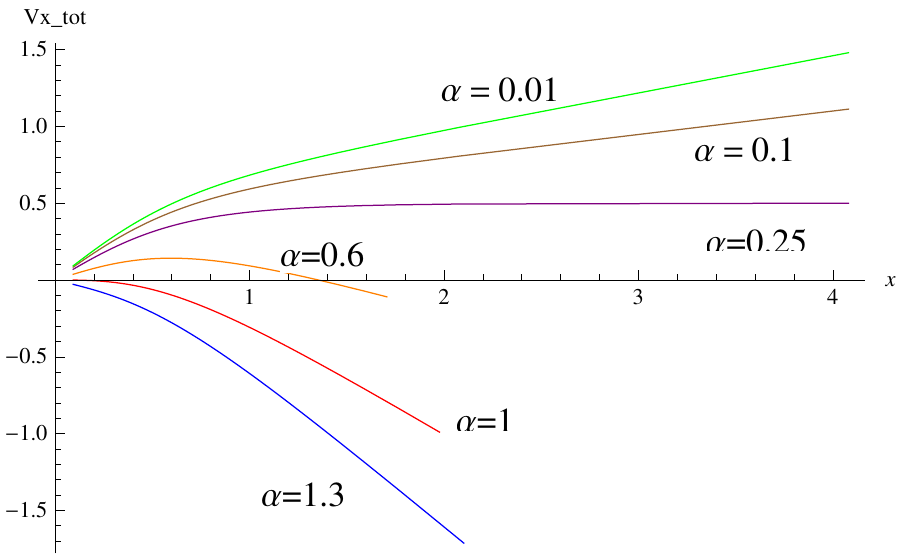}
  \captionof{figure}{AdS soliton} \label{fig:soliton1}
  \label{fig:test15}
\end{minipage}%
\begin{minipage}{.5\textwidth}
  \centering
  \includegraphics[width=.9\linewidth]{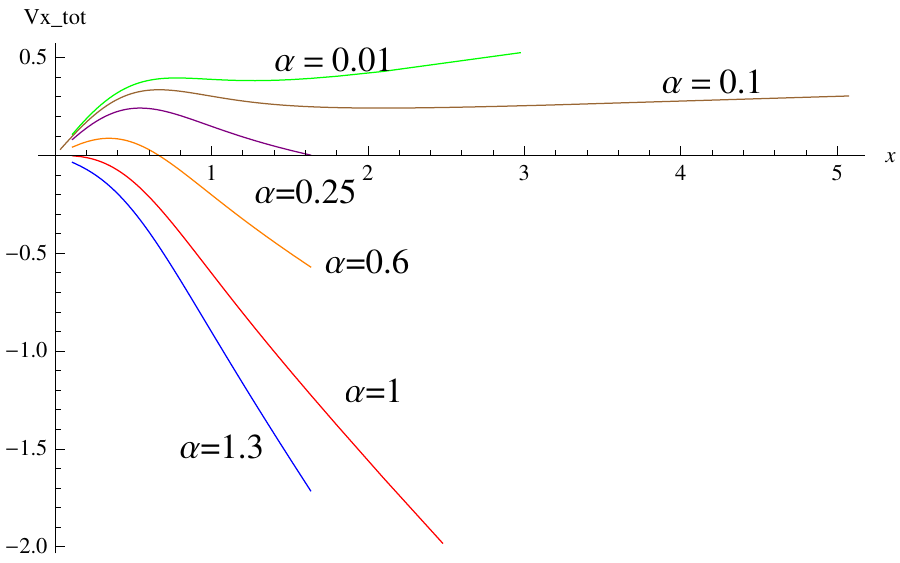}
  \captionof{figure}{Witten QCD}
  \label{fig:test23}
\end{minipage}
\end{figure}

In \cite{Sato:2013hyw} the authors demonstrated some generals features in the phase diagram of Schwinger effect in all confining geometries and then in \cite{Sato:2013dwa}, they have studied the ``AdS soliton geometry" as a special case. In these papers, the authors studied the potential of a confining background and then they have plotted the total potential, $V_{\text{tot}} (x)$ versus the distance between the quark and anti quark $x$. Their setup and the world sheet configuration for the quark and anti quark potential is also shown, in Figures \ref{fig:test13a} and \ref{test14b}.
To compare with our diagrams, their plot of the phase diagram for the AdS soliton geometry \cite{Sato:2013hyw} is re-produced in Figure \ref{fig:soliton1}.

Now in the following sections, we do a similar calculation for our class of supergravity confining geometries and then numerically we find the phase diagrams.

\subsection{Witten-QCD}

First for the Witten-QCD model, if we assume that a probe D3-brane is located at $u=u_0$ and then we assume the following ansatz for the Wilson loop,

\begin{gather}
x^0=\tau, \ \ \ \ x^1=\sigma, \ \ \ u=u(\sigma), \ \ \ \theta=\theta(\sigma),
\end{gather}
the NG action would be
\begin{gather}
\mathcal{L}=\sqrt{-g}= \sqrt{ \left(\frac{u}{R}\right)^3 \left(1+\frac{4 R^3}{9 u_t}  {\theta'} ^2 \left(1-\frac{{u_t}^3}{u^3}\right)+\frac{{u'}^2}{1-\frac{u_t^3}{u^3}}\left (\frac{R}{u}\right)^3 \right)   }.
\end{gather}
As the Lagrangian does not depend on $\sigma$, the following Hamiltonians are conserved,
\begin{gather}
H_u=\frac{\partial \mathcal{L}}{\partial(\partial_\sigma u)}\partial_\sigma u -\mathcal{L}, \ \ \ \ H_\theta=\frac{\partial \mathcal{L}}{\partial(\partial_\sigma \theta)}\partial_\sigma \theta -\mathcal{L}.
\end{gather}
Then there should exist two different minimal surfaces that satisfy the following relations,
\begin{gather}
\frac{du}{d\sigma}=0, \ \ \text{at} \ \ u=u_c \ \ (u_t<u_c<u_0 ),  \ \ \ \ \ \ \frac{d\theta}{d\sigma}=0, \ \ \text{at} \ \ \theta=\theta_c \ \ (\theta_t<\theta_c<\theta_0 ).
\end{gather}
However by assuming $\theta'=0$, we consider a rectangular Wilson loop which makes the calculation simpler. So
\begin{gather}
\mathcal{L}=\sqrt{\frac{1}{1-\frac{u_t^3}{u^3} } \left(\frac{du}{d\sigma}\right)^2+\frac{u^3}{R^3} },
\end{gather} 
and then from the conservation of $H_u$ we would get,
\begin{gather}
\frac{du}{d\sigma}=\frac{1}{R^{\frac{3}{2}}} \sqrt{(u^3-{u_t}^3) \left( \frac{u^3}{u_c^3}-1  \right) }.
\end{gather}
By integrating the above equation one could get the length of the string between the quark and anti quark in WQCD as
\begin{gather}
x=2  R^{\frac{3}{2}} \int_{u_c}^{u_0} \frac{ du }{ \sqrt{ (u^3-u_t^3)\left( (\frac{u}{u_c})^3-1  \right) }}.
\end{gather}
Now by defining the following dimensionless quantities,
\begin{gather}
y=\frac{u}{u_c}, \ \ \ \ a=\frac{u_c}{u_0}, \ \ \ \ \ b=\frac{u_t}{u_0},
\end{gather}
one can simplify $x$ as
\begin{gather}
x= \frac{ 2 R^{\frac{3}{2}}}{(u_0 a)^{\frac{1}{2}} } \int_{1}^{\frac{1}{a}} \frac{ dy }{ \sqrt{ (y^3-1)\Big(y^3-(\frac{b}{a})^3 \Big) }}.
\end{gather}
Then the sum of the potential and static energy would be
\begin{gather}
V_{PE+SE}=2T_F\int_0^{\frac{x}{2}} d\sigma \mathcal{L}=2T_f u_0 a \int_1^{\frac{1}{a}} dy \frac{y^3}{\sqrt{(y^3-1)\Big (y^3-\frac{b^3}{a^3} \Big )}}.
\end{gather}
For the large $x$ limit, ($a\to b$), the sum of the potential and static energy is
\begin{gather}
V_{PE+SE}=T_F\frac{({u_0 b}) ^{\frac{3}{2} }}{R^{\frac{3}{2}}}x+2 T_F u_0 b  \left( \frac{1}{b}-1 \right).
\end{gather}
So from the first term which is the quark and anti-quark potential, we can read the confining string tension as
\begin{gather}
\sigma_{\text{st}}= T_F {\left(\frac{u_t}{R} \right)^{ \frac{3}{2} }}.
\end{gather}
This matches with the result coming from the relation $\sigma_{st}=\frac{g{(u_t)}}{2\pi \alpha'}$.
Also the second term gives the static mass of the quark and anti-quark,
\begin{gather}
2T_F (u_0-u_t)=2m_W.
\end{gather}

Then from the DBI action, one can read the critical electric field for this geometry as
\begin{gather}
E_c= T_F  \Big( \frac{u_0}{R} \Big)^{\frac{3}{2}}.
\end{gather}
Now we can define the dimensionless parameter $\alpha=\frac{E}{E_c}$. So the total potential energy would be
\begin{gather}
V_{\text{tot}}=V_{\text{PE}+\text{SE}}-Ex=\nonumber\\
2T_Fu_0 a \int_1^{\frac{1}{a}} dy \frac{y^3}{ \sqrt{(y^3-1) (y^3-\frac{b^3}{a^3} )}}-\frac{2T_F u_0 \alpha}{a^{\frac{1}{2}}} \int_1^{\frac{1}{a}} \frac{dy}{ \sqrt{(y^3-1) (y^3-\frac{b^3}{a^3} )}}
\end{gather}

\begin{figure}[h!] \label{fig:wQCD}
  \centering
    \includegraphics[width=0.6\textwidth]{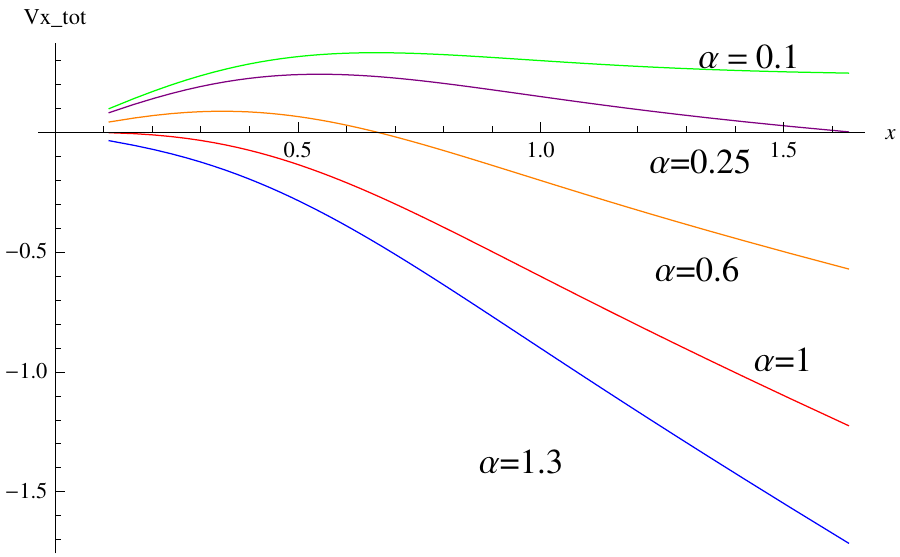}
       \caption{The plot of total potential versus $x$ for Witten QCD model when $b=0.5$ and $2T_F=u_0=1$ . }
\end{figure}

By comparing the diagrams below, one can see that the form of plots are similar with slight differences in small $x$ limit.  One can see that in both of them, there exist three phases, one with stable potentials  and no pair creation, one with exponentially suppressed potentials with tunneling pair creation and one with catastrophically unstable potentials with exponential pair creation. 

As it has been demonstrated analytically for a general background in \cite{Sato:2013hyw} this behavior is universal in all confining geometries. However there are still some minor differences between the phase diagrams of different confining backgrounds which here we aim to detect and then compare with the phase diagrams of the entanglement entropy of these backgrounds. 

From these plots one can see that in Witten QCD diagram, for $\alpha=0.01$ there is no zero other than the origin and no Schwinger effect can occur. For $\alpha=0.1$ the potential becomes flat.  For $\alpha=0.25$ or 0.6, there is a barrier in the potential which as it can be seen is different from the AdS soliton case with a bigger curvature. In this phase the Schwinger pair creation can only occur by tunneling through this barrier and the rate of pair creation is strongly suppressed. For larger $E$ as $\alpha=1.3$ for instance, the potential becomes catastrophically unstable and the Schwinger effect would occur and therefore a current can be created. In this phase the probability of pair creation would not be any more exponentially suppressed. 
The curvatures are bigger for the Witten QCD case and a small bump can be seen at around $x=0.5$ which is not present for the AdS soliton case. Also comparing all the diagrams one can see that in Witten-QCD and KT the phase transition happens faster and in a more dramatic way relative to the other backgrounds.

\subsection{Maldacena-Nunez}

\begin{figure}[h!] 
  \centering
    \includegraphics[width=0.3\textwidth]{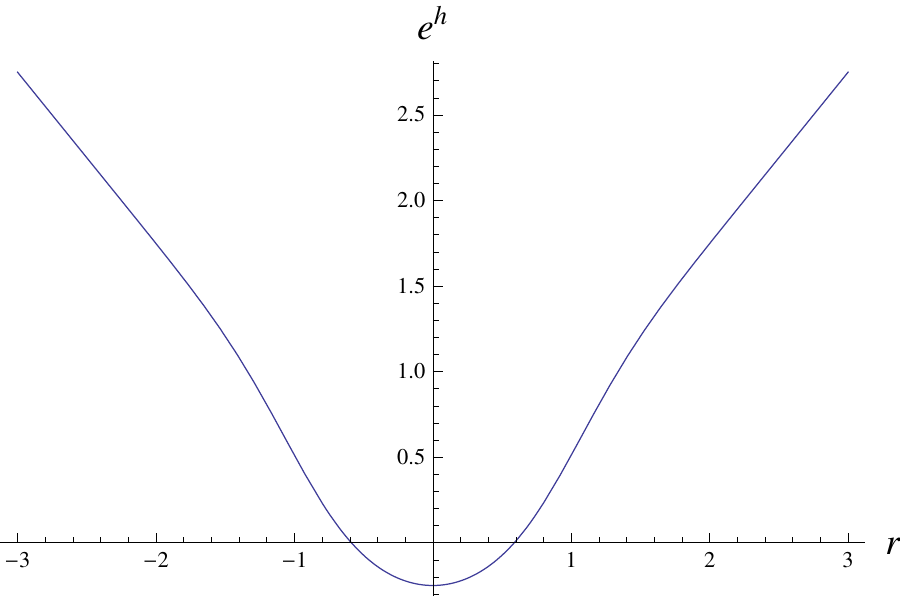}
       \caption{The Plot of $e^h$ versus $r$ . }
       \label{fig:roots1}
\end{figure}

Now we repeat this calculation for the Maldacena-Nunez background.
We assume $x^0=\tau, \ x^1=\sigma, \ r=r(\sigma)$, so the determinant of the induced metric and therefore the Lagrangian is $ \mathcal{L}=e^\phi \sqrt{r'^2+1}$.
Since
\begin{gather}
\frac{\partial \mathcal{L}}{\partial(\partial_\sigma r)} \partial_\sigma r-\mathcal{L},
\end{gather}
is a constant, we would get
\begin{gather}
r'=\sqrt{\frac{e^{2\phi}}{e^{2\phi({r_c})}}-1}.
\end{gather}
So, $\mathcal{L}=\frac{e^{2\phi}}{e^{\phi_c}}$. Also from the DBI action one can find the critical electric field as
\begin{gather}
E_c=T_F \frac{e^{2\phi({r_0})} }{e^{\phi({r_c})}}.
\end{gather}
We assume that the probe brane is located at $r=r_c=1$, and for the sake of similarity to the previous calculations we define $a=\frac{r_c}{r_0}$, therefore $r_0=\frac{1}{a}$.
One should also notice that the MN geometry ends when $e^\phi$ becomes undefined which happens at the roots of $e^h=r \coth(2r) -\frac{r^2}{\sinh[2r^2]}-\frac{1}{4}$ where its plot is shown in Figure \ref{fig:roots1} .

\begin{figure}[h!] 
  \centering
    \includegraphics[width=0.6\textwidth]{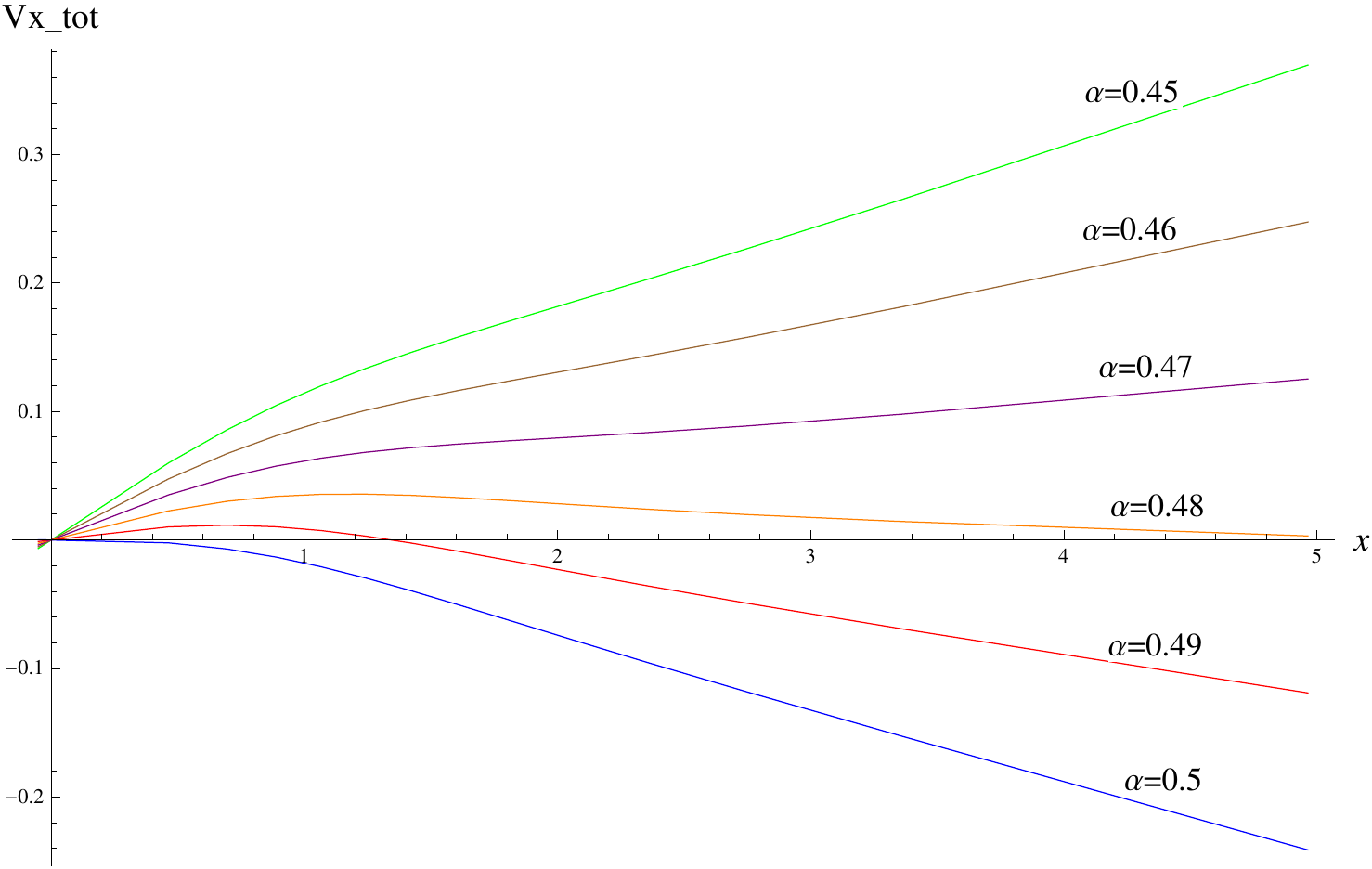}
       \caption{The plot of total potential versus $x$ for the Maldacena-Nunez model. Here $\phi_0=0$, $r_0=T_F=1$. }
  \label{fig:MNa}
\end{figure}

One can see that the positive root is approximately located at $r_t=0.6$ where the geometry ends.
Now by defining 
\begin{gather}
x=2\int_{r_c}^ {r_0} \frac{dr}{ \sqrt{\frac{e^{2\phi}}{ e^{2\phi_c}} -1 }},
\end{gather}
and the total potential as,
\begin{gather}
V_{\text{tot}}=2T_F \int_{r_c}^{r_0} \frac{e^{2\phi} }{e^{2\phi_c}} \frac{ dr}{\sqrt{\frac{e^{2\phi} }{e^{2\phi_c}  }-1 } } -2E_c \alpha  \int_{r_c}^{r_0} \frac{dr}{\sqrt{\frac{e^{2\phi} }{e^{2 \phi_c} }-1 }},
\end{gather}
 one can find the phase diagram in Figure \ref{fig:MNa}. 
 
By expanding the potential $V_{\text{PE}+\text{SE}}$ for large $x$, i.e, $r \to r_c$, one can find the string tension and quark mass as,
\begin{gather}
\sigma_{st}=e^{\phi_c} T_F, \ \ \ \ \ \  2m_W=2T_F e^{\phi_c} (r_0-r_t).
\end{gather}
This result for the string tension matches with other relation from $\frac{g(r_t)}{2\pi \alpha'}$.
One can see that still there exists three different phases. For $\alpha$ till  around 0.47 no Schwinger effect would happen. For $\alpha$ between 0.47 and 0.48 Schwinger effect would occur as a tunneling process, and for $\alpha$ larger than 0.48, the potential becomes unstable. 

\subsection{Klebanov-Strassler}

\begin{figure}[h!]
  \centering
    \includegraphics[width=0.6\textwidth]{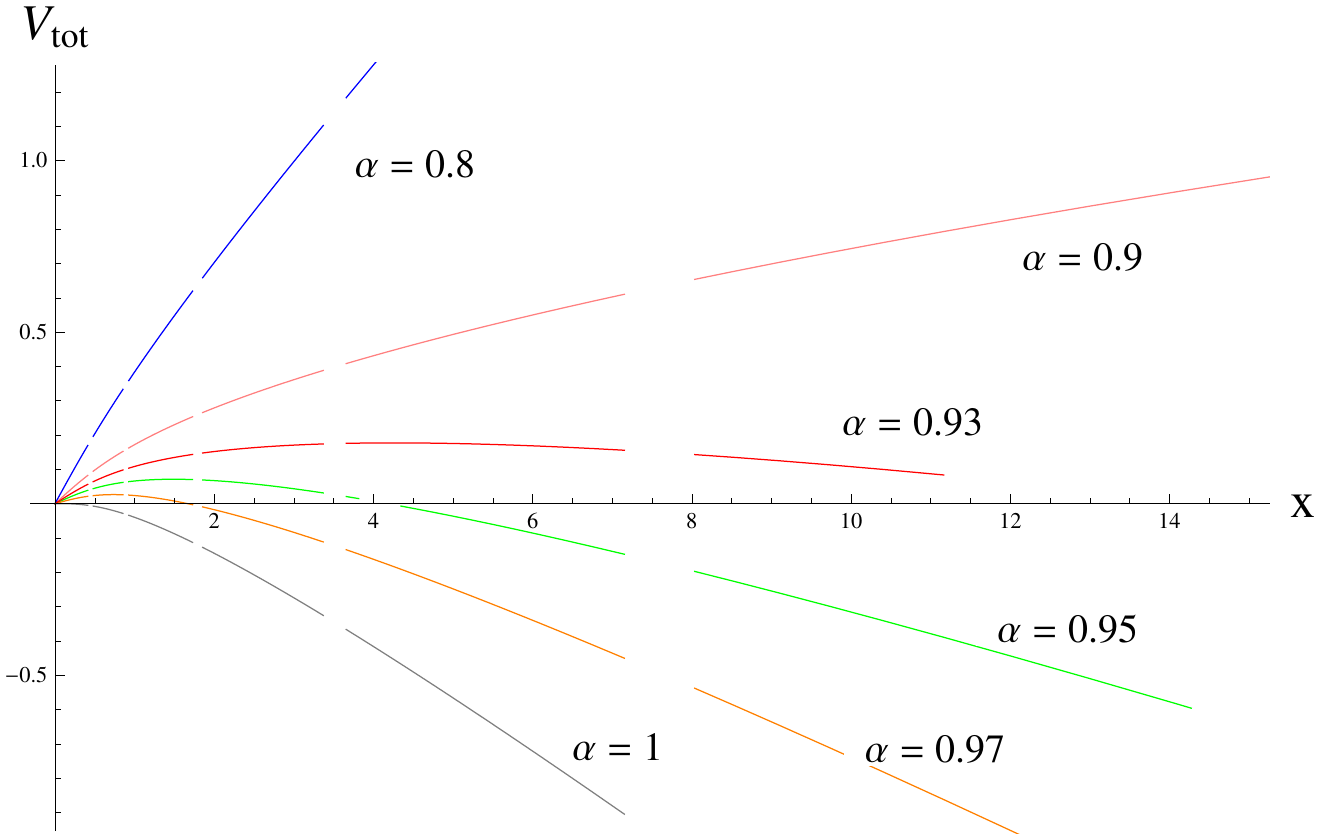}
       \caption{The plot of total potential versus $x$ for the Klebanov-Strassler model, assuming $\tau_0=\beta=T_F=\epsilon=1$ . }
  \label{fig:KSa}
\end{figure}

Now for the KS background we assume $x^0=t, \ x^1=\sigma,\  \tau=\tau(\sigma)$.
So the components of the induced metric are
\begin{gather}
\gamma_{tt}=-h^{-\frac{1}{2}}(\tau), \ \ \ \ \ \gamma_{\sigma \sigma}=h^{-\frac{1}{2}}(\tau)+\frac{\tau'^2 h^{\frac{1}{2}}(\tau) \epsilon^{\frac{4}{3}}}{6 K^2(\tau) }.
\end{gather}
Therefore
\begin{gather}
\mathcal{L}= \sqrt{h^{-1}(\tau)+\frac{\tau'^2 \epsilon^{\frac{4}{3}}}{6K^2(\tau) } }.
\end{gather}
Again knowing that $\frac{\partial \mathcal{L}}{\partial(\tau')}-\mathcal{L}$ is constant, one can derive $\tau'$ as
\begin{gather}
\tau'=\frac{\sqrt{6}} {\epsilon^{\frac{2}{3}}} \frac{K(\tau)}{h(\tau)} \sqrt{h(\tau_c)-h(\tau) }.
\end{gather}
Also the critical electric field is 
\begin{gather}
E_x= T_F h^{-\frac{1}{2}} (\tau_0).
\end{gather}
At $\tau=\tau_c$ one would have $\tau'_c=0$. We assume $\tau_0=1$, so one can find
\begin{gather}
x=2 \int_1^{\frac{1}{a}} \frac{\epsilon^{\frac{2}{3}} h(ya) }{\sqrt{6} K(ya) } \frac{1}{ \sqrt{h(a) -h(ya)} },
\end{gather}
and the total potential as
\begin{gather}
V_{tot}=\frac{2 \epsilon T_F}{\sqrt{6}} \int_1^{\frac{1}{a}} \frac{ \sqrt{h(a)} }{K(ya)} \frac{1}{\sqrt{h(a)-h(ya) } }\nonumber\\
-\frac{2\epsilon^{\frac{2}{3}} T_F \alpha}{\sqrt{6} } \int_1^{ \frac{1}{a}} \frac{h(ya) }{K(ya) } \frac{h^{-\frac{1}{2}} (1)}{\sqrt{h(a)-h(ya) }}.
\end{gather}
The plot of KS phases for different $\alpha$ is shown in Figure \ref{fig:KSa}.

The numerical calculation of this case would take more time to finish but the final phase diagram in general is similar to the three other confining geometries. The part that are missing in the plot are due to some singularities in the numerical calculations of the integrals. Again three phases for different $\alpha$ can be detected and the phase transitions are mild.

\subsection{Klebanov-Tseytlin}
The calculation can be repeated for the KT background.
Taking  $x^0=t, \ x^1=\sigma,\  r=r(\sigma)$, the components of the induced metric are
$\gamma_{tt}=h^{-\frac{1}{2}}, \ \gamma_{\sigma \sigma}=-(h^{-\frac{1}{2}}+r'^2 h^{\frac{1}{2}} )$ and 
 so $\mathcal{L}= \sqrt{h^{-1} +r'^2}$.
 
Again using the conservation of the Hamiltonian and knowing  that at $r=r_c$, $r'=0$, then we find
\begin{gather}
r'=\frac{\sqrt{h(r_c)-h(r)} }{h(r)}.
\end{gather}

After taking the integral of this relation and by defining $y=\frac{r}{r_c} $ , $a=\frac{r_c}{r_0}$ and $b=\frac{r_s}{r_0}$, one can find

\begin{gather}
x=9 \sqrt{2} M \epsilon ^2 \sqrt{g_s } \int_{1} ^{\frac{1}{a}}\frac{dy}{y^4 \sqrt{\ln \left(\frac{a}{b}\right) -\frac{1}{y^4} \ln \left(\frac{ya}{b} \right) } } ,\nonumber\\
V_{PE+SE}=2T_F\int_{r_0}^{r_s} \mathcal{L} d\sigma=2T_F a r_0 \sqrt{\ln \frac{a}{b}} \int_1^{\frac{1}{a}} \frac{dy}{\sqrt{\ln \frac{a}{b}-\frac{1}{y^4}\ln \frac{ay}{b}  } }.
\end{gather}
From the DBI action the critical electric field is
\begin{gather}
E_{c}=T_F h_0^{-\frac{1}{2}}= T_F \frac{r_0^2}{L^2} \left(\ln \frac{r_0}{r_s} \right)^{-\frac{1}{2}}.
\end{gather}
By defining $\alpha=\frac{E}{E_{c}}$, the total potential is, 
\begin{gather}
V_{\text{tot}}=2T_F a r_0 \sqrt{\ln \frac{a}{b}} \int_1^{\frac{1}{a}} \frac{dy}{\sqrt{\ln \frac{a}{b}-\frac{1}{y^4}\ln \frac{ay}{b}  } }- \frac{2\alpha T_F r_0^2}{\sqrt{ \ln\frac{1}{b}} }  \int_{1} ^{\frac{1}{a}}\frac{dy}{y^4 \sqrt{\ln \frac{a}{b}-\frac{1}{y^4} \ln \frac{ya}{b}  } }.
\end{gather}

\begin{figure}[h!]
  \centering
    \includegraphics[width=0.6\textwidth]{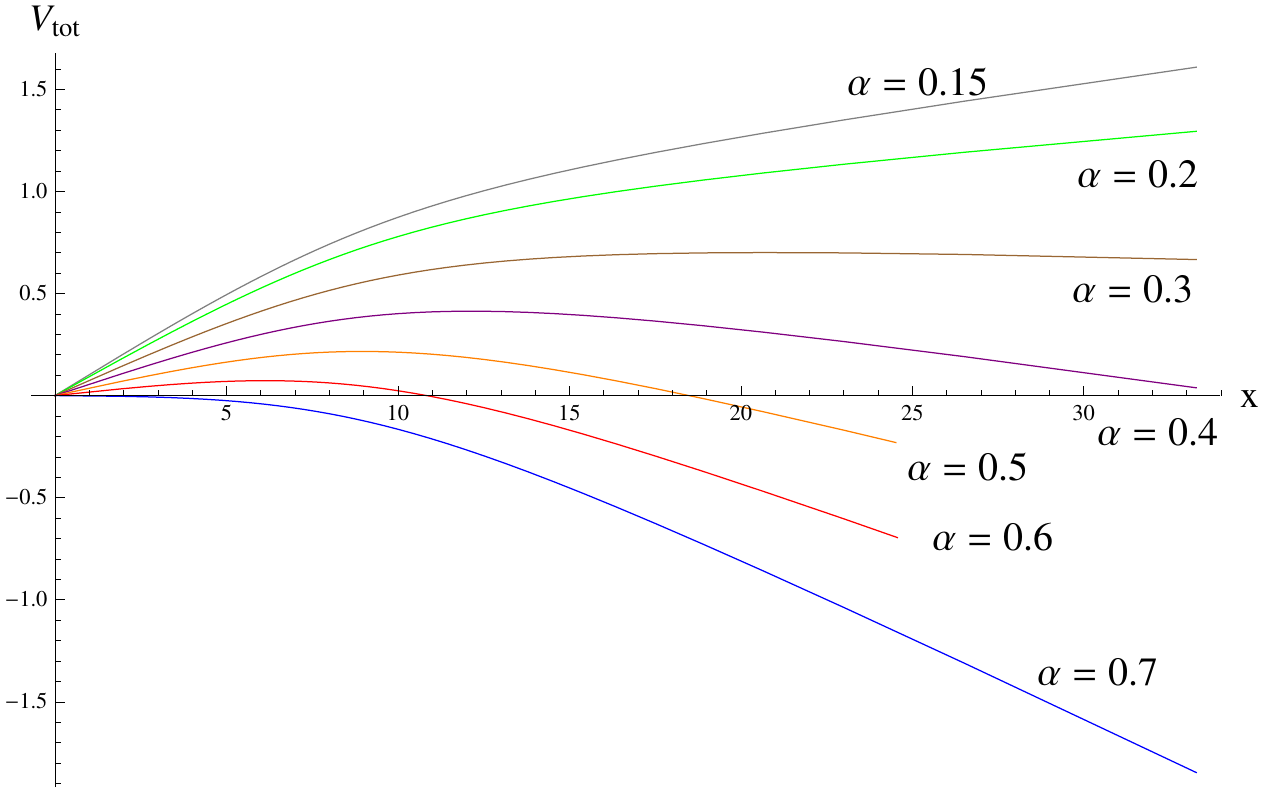}
       \caption{The plot of total potential versus $x$ for the Klebanov-Tseytlin model for $b=0.5$ and $r_0=M=T_F=1$ . }
        \label{fig:KTa}
\end{figure}

Now by simplifying this relation by assuming $b=0.4$ and  $r_0=M=T_F=1$, one can find the plot of $V_{\text{tot}}$ versus $x$ which is shown in Figure \ref{fig:KTa}.

The Klebanov-Tseytlin model is a limit of Klebanov-Strassler and there is no wonder that the phase diagram would be very similar. Again three phases can be seen for different $\alpha$. However in these two cases the exact numerical values of $\alpha$ does not correspond to each other,  since we didn't match the numerical constants.

\subsection{Klebanov-Witten}

\begin{figure}[h!]
  \centering
    \includegraphics[width=0.6\textwidth]{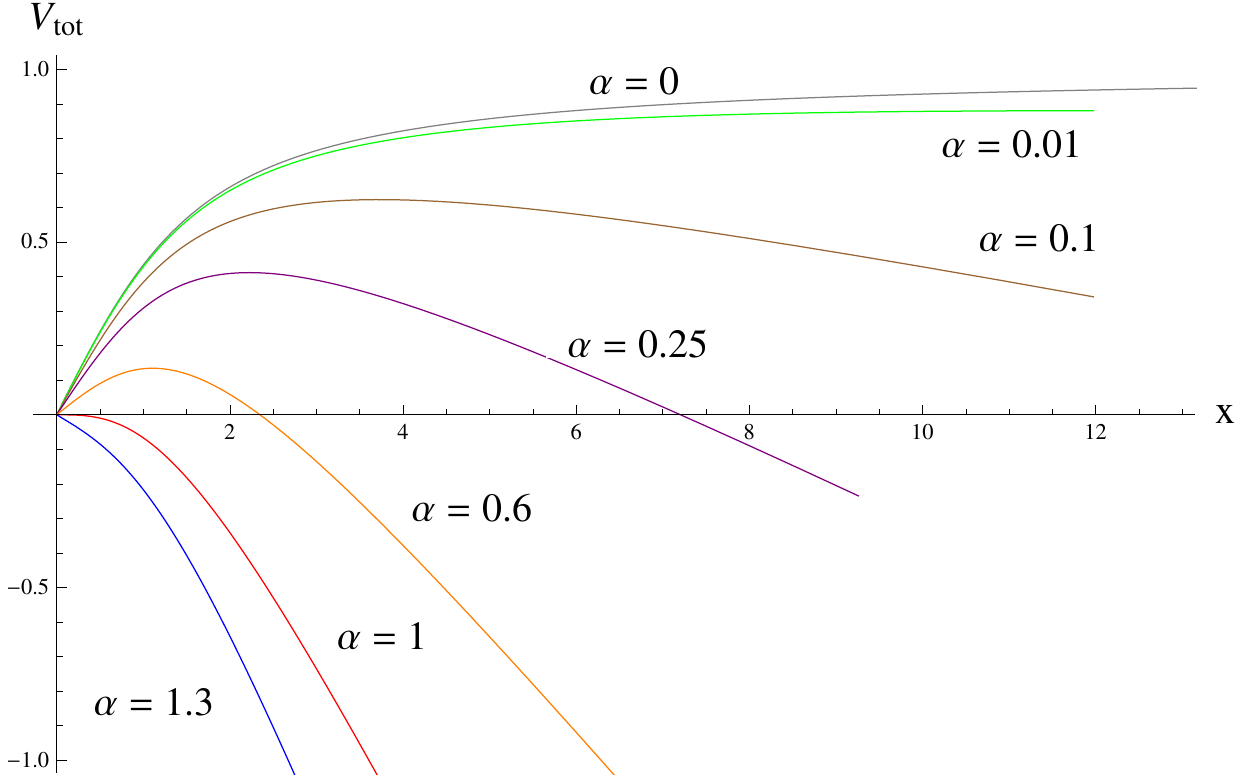}
       \caption{The Schwinger phases of $\text{AdS}_5$ and Klebanov-Witten \cite{Sato:2013iua}. }
        \label{fig:11a}
\end{figure}

Now for comparing our results of the phase diagrams of confining geometries with the conformal backgrounds, we also study the Klebanov-Witten geometry.  The phase diagram would be exactly identical to the phase diagram of AdS case which has been studied in \cite{Sato:2013iua}.
Assuming $ x^0=t, \ x^1=\sigma, \ r=r(\sigma)$, 
where $h=\frac{L^4}{r^4}$. Then $\mathcal{L}= \sqrt{h^{-1} +r'^2}$, and as the Hamiltonian is conserved, one would get,
\begin{gather}
r'=\frac{r^2}{L^2}\sqrt{ \frac{r^4}{r_c^4} -1},
\end{gather}
where $\frac{dr}{d\sigma}=0$ at $r=r_c$. So $\mathcal{L}=\frac{r^4}{L^2 r_c^2}$ and
from the DBI action the critical electric field is  $E_x=\frac{T_F r_0^2} {L^2}$. By integrating $r'$ and by defining $y=\frac{r}{r_c}$, $a=\frac{r_c}{r_0}$ and $b=\frac{r_t}{r_0}$, ($r_t=0$ here) the distance between the quark and anti-quark and the potential are
\begin{gather}
x=\frac{2L^2}{r_0 a} \int_1^{\frac{1}{a}} \frac{dy}{y^2 \sqrt{y^4-1} }, \\
V_{PE+SE}= 2T_F r_0 a  \int_1 ^{\frac{1}{a}} \frac{y^2 dy}{ \sqrt{y^4-1} }.
\end{gather}
So for large $x$, as $a\to b=0$, the potential is ``zero" which is a feature of non-confining, conformal backgrounds. Also, $2m_W=2T_F r_0 (1-a)$ and for $a=b=0$, it gives $2m_W=2T_F r_0$. 
Then
\begin{gather}
V_{tot}=2T_F r_0 a \int_1 ^{\frac{1}{a}} \frac{y^2 dy}{\sqrt{y^4-1} }-\frac{2T_F r_0 \alpha }{a}  \int_1^{\frac{1}{a}} \frac{dy}{y^2 \sqrt{y^4-1} }.
\end{gather}
The phase diagram of KW and AdS are exactly similar and is shown in Figure \ref{fig:11a}.

Unlike the other four confining geometries, as the KW and AdS are conformal geometries, there are only two phases here. Even for a very small $\alpha$ or electric fields, there is a zero at larger $x$ and so the pair creation would happen by a tunneling process. The other phase is the unstable one for a larger $\alpha$.

\section{Entanglement entropy of a strip in confining geometries}
Now we would like to study the phase diagrams of entanglement entropy in these confining geometries  and then compare the phase transition in EE with the Schwinger effect and compare the diagrams of different geometries. So one way is to use the method in \cite{Hubeny:2014zna} for calculating the entanglement entropy of accelerating quark and anti-quark. But as we have mentioned in section 2, one first needs to solve several PDE equations analytically. Other methods could be using the ideas in \cite{Caputa:2014vaa} or \cite{Wong:2013gua},  to study the entanglement entropy of local operators or localizes excited states in each background. But here we follow the calculation of \cite{Kol:2014nqa} and \cite{Klebanov:2007ws} to find the entanglement entropy of a strip in each confining geometries. Also the entanglement entropy of multiple strips can be calculated similar to \cite{Ben-Ami:2014gsa}.

So as in \cite{Klebanov:2007ws},  based on KKM, the generalization of Ryu-Takayanagi conjecture for the non-conformal theories would be
\begin{gather}
S=\frac{1}{G_N^{(10)}} \int_\gamma d^8 \sigma e^{-2\phi} \sqrt{G_{{ind}}^{(8)}} .
\end{gather}
The authors showed that the entanglement entropy can be found by minimizing this action over all surfaces that ends on the boundary of the entangling surface and there are actually two solutions, where one of them corresponds to a disconnected region and the other one is a connected surface. In\cite{Wong:2013gua}, the authors showed that always, only one of the two possible configuration would dominate. 

So if one writes the gravitational background in the following form of,
\begin{gather}
ds^2=\alpha(\rho) [ \beta(\rho) d\rho^2+dx^\mu dx_\mu]+g_{ij} d\theta^i d\theta^j, \ \ \ \  (\mu=0,1,...d),\ \ (i=d=2,...,9),
\end{gather}
then the volume of the internal manifold is $V_{\text{int}}=\int d\vec{\theta} \sqrt{\text{set} [g_{ij}]}$, and one can define another function $H(\rho)$ as
\begin{gather}
H(\rho)=e^{-4\phi} V_{\text{int}}^2 \alpha^d.
\end{gather}
This function for confining geometries is a monotonically increasing function while $\beta(\rho)$ is a monotonically decreasing function.
Using the KKM equation \cite{Klebanov:2007ws},\cite{Kol:2014nqa} the EE for the connected solution is
\begin{gather}
S_C(\rho_0)=\frac{V_{d-1}}{2 G_N^{(10)}} \int_{\rho_0} ^\infty d\rho \sqrt { \frac{\beta(\rho) H(\rho)} {1-\frac{H(\rho_0)}{H(\rho)} } },
\end{gather}
and the EE of the disconnected solution is given by
\begin{gather}
S_D(\rho_0)= \frac{V_{d-1}}{2G_N^{(10)}} \int_{\rho_\Lambda}^\infty d\rho \sqrt{ \beta(\rho) H(\rho) },
\end{gather}
and the length of the line segment of the connected solution is
\begin{gather}
L(\rho_0)=2\int_{\rho_0}^\infty d\rho \sqrt{ \frac{\beta(\rho)}{\frac{H(\rho)}{H(\rho_0)}-1 } }.
\end{gather}
The difference of the connected and disconnected solution would be finite, so $S$ is defined as
\begin{gather}
S(\rho_0)=\frac{2G_N^{(10)}}{V_{d-1}} (S_C-S_D)=\int_{\rho_0} ^\infty d\rho \sqrt { \frac{\beta(\rho) H(\rho)} {1-\frac{H(\rho_0)}{H(\rho)} } }-\int_{\rho_\Lambda}^\infty d\rho \sqrt{ \beta(\rho) H(\rho) }.
\end{gather}

Now we study $L$ and $S$ for our specific geometries.

\subsection{Witten-QCD background}

For the case of Witten QCD background similar to \cite{Kol:2014nqa}, one can define the functions $\alpha(u), \beta(u)$ and $H(u)$ as
\begin{gather}
\alpha(u)=\left(\frac{u}{R}\right)^{\frac{3}{2}}, \ \ \ \ \ \beta(\rho)=\left(\frac{R}{u}\right)^3 \frac{1}{f(u)},\ \ \ \ \ \ H(u)=\left( \frac{8\pi^2}{3}\right)^2 \frac{4R^6 u^5 f(u)}{9 u_0 g_s^4} ,
\end{gather}
where $f(u)=1-\frac{u_t^3}{u^3}$. From the functions of $\beta$ and $H$, it can be seen that as it has been suspected, $\beta(u)$ is monotonically decreasing, $H$ is monotonically increasing, $H$ shrinks to zero at $u=0$ and $\beta(u)$ diverges at $u=u_t=1$.

\begin{figure}[h!]
\centering
\begin{minipage}{.55\textwidth}
  \centering
  \includegraphics[width=.9\linewidth]{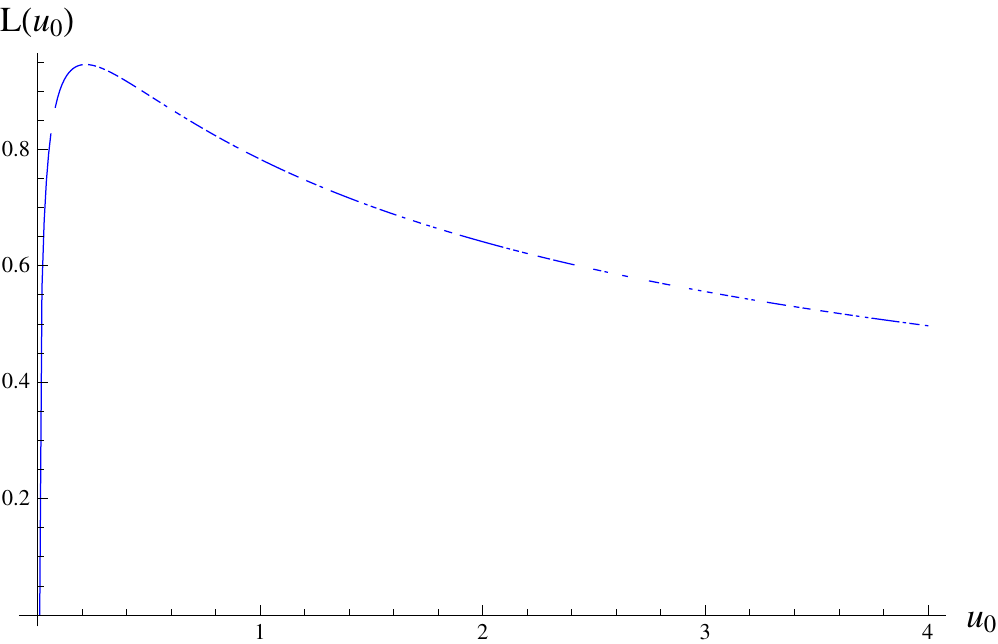}
  \captionof{figure}{Plot of $L(u_0)$ v.s $u_0$ for WQCD model. }
  \label{fig:WQCDL}
\end{minipage}%
\begin{minipage}{.55\textwidth}
  \centering
  \includegraphics[width=.9\linewidth]{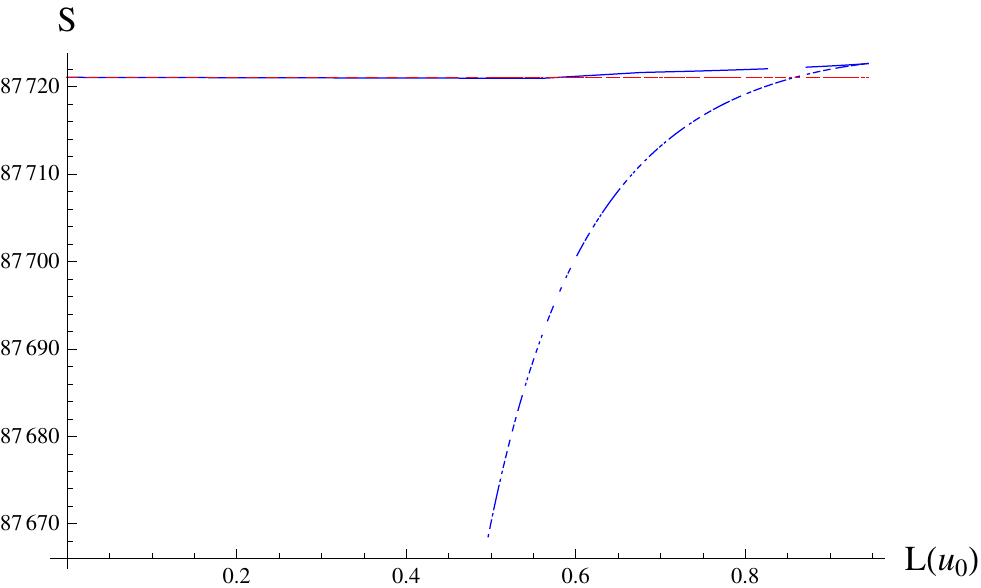}
  \captionof{figure}{Plot of $S$ vs. $u_0$ for WQCD model. The blue line is the connected solution and the dashed red line is the disconnected solution. }
  \label{fig:WQCDS}
\end{minipage}
\end{figure}

The EE for the connected and disconnected solution are respectively 
\begin{gather}
S_{C}(u_0)=\frac{V_2}{ {G_N}^{(10)}} \frac{8\pi^2 R^{\frac{9}{2}}}{9g_s^2 \sqrt{u_t} }  \int _{u_0}^{\infty} du \sqrt{ \frac{u^2}{1-\frac{u_0^5}{u^5} \frac{f(u_0)}{f(u)}  } },\nonumber\\
S_D(u_0)=\frac{V_2}{G_N^{10}} \frac{8\pi^2 R^{\frac{9}{2}} }{9 g_s^2 \sqrt{u_t}} \int_{u_\Lambda}^\infty u du.
 \end{gather}
 Also the length of the line segment of the connected solution is
\begin{gather}
 L(u_0)=2\int_{u_0}^\infty du \sqrt { \frac{(\frac{R}{u})^3 \frac{1}{f(u)} }{\frac{f(u) u^5}{f(u_0) u_0^5} -1} }.
 \end{gather}
 Their plots are shown in Figures \ref{fig:WQCDL} and \ref{fig:WQCDS}.

Here we take $u_t=u_{\Lambda}=1$. 
Due to the peak in the plot of $L$ and also the butterfly shape and the double valuedness in the plot of $S$, one can deduce that a phase transition and therefore a confinement phase exist. This phase transition and the shape of the peak are similar to the instability of $V$ and the phase transition in the previous section. One can see that the phase transition is more dramatic relative to the other geometries. This dramatic phase transition in smaller $x$ were also seen in the Schwinger phase transition in WQCD relative to the other geometries. So there might be a deeper relationship between these two quantities.

\subsection{Klebanov-Tseytlin}
For the KT case, the volume of the internal part, $ds_{T^{1,1}}$, is $\frac{16 \pi^3}{27}$. So for the KT background the functions would be
\begin{gather}
\alpha(r)=h(r)^{-\frac{1}{2}},  \ \ \ \ \  \beta(r)=h(r), \ \ \ \ \  V_{\text{int}}=\frac{16 \pi^3}{27} r^5 h(r)^{\frac{5}{4}}, \ \ \ \ \ 
H(r) = \left( \frac{16}{27} \right)^2 \pi^6 r^{10} h(r).
\end{gather}

\begin{figure}[h!]
\centering
\begin{minipage}{.5\textwidth}
  \centering
  \includegraphics[width=.9\linewidth]{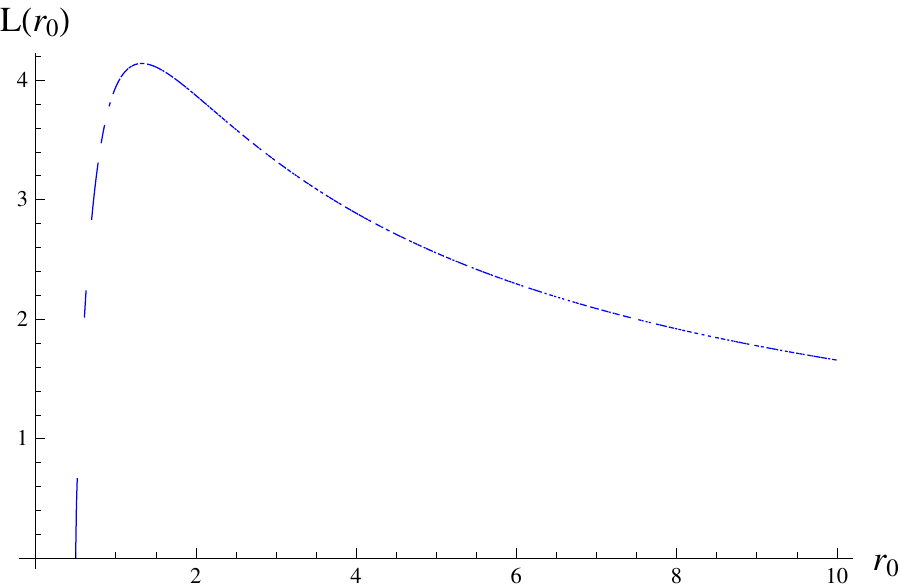}
  \captionof{figure}{Plot of $L(r_0)$ vs $r_0$ for KT, $r_s=0.5$. }
  \label{fig:KTL2}
\end{minipage}%
\begin{minipage}{.5\textwidth}
  \centering
  \includegraphics[width=.9\linewidth]{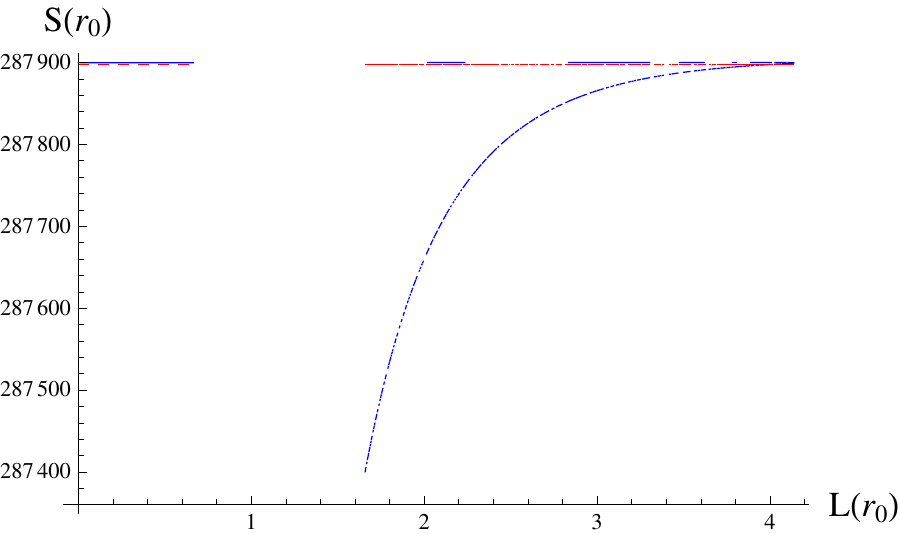}
  \captionof{figure}{Plot of $S(r_0)$ vs $L$ for KT, $r_s=0.5$. }
  \label{fig:KTSS2}
\end{minipage}
\end{figure}

Again $\beta$ and $H$ show the expected monotonically decreasing and increasing behaviors. Then the length of the connected region and the EE are

\begin{gather}
L(r_0)= 9\sqrt{2}   M \sqrt{g_s} \epsilon^2  \int_{r_0}^\infty dr \frac{ \sqrt{\frac{r}{r_s}} }{r^2 \sqrt{\frac{r^6}{{r_0}^6} \frac{\ln {\frac{r}{r_s} } }{\ln \frac{r_0}{r_s} } -1} }  ,\nonumber\\
S_C(r_0)=\frac{12 V_2 \pi^3 M^2 g_s  \epsilon^4}{G_N^{(10)}} \int_{r_0}^\infty dr \frac{r \ln \frac{r}{r_s} }{\sqrt{1-\frac{r_0^6}{r^6}\frac{\ln \frac{r_0}{r_s} }{\ln \frac{r}{r_s}}  } },\nonumber\\
S_D(r_0)=\frac{12V_2 \pi^3  M^2  g_s \epsilon^4 }{G_N ^{ (10)}} \int_{r_\Lambda} ^\infty dr \ r \ln \frac{r}{r_s}.
\end{gather}
The plots for the $M=\epsilon=g_s=1$ and $r_s=0.5$ is shown bellow in Figures \ref{fig:KTL2} and \ref{fig:KTSS2}.

\begin{figure}[h!]
\centering
\begin{minipage}{.5\textwidth}
  \centering
  \includegraphics[width=.9\linewidth]{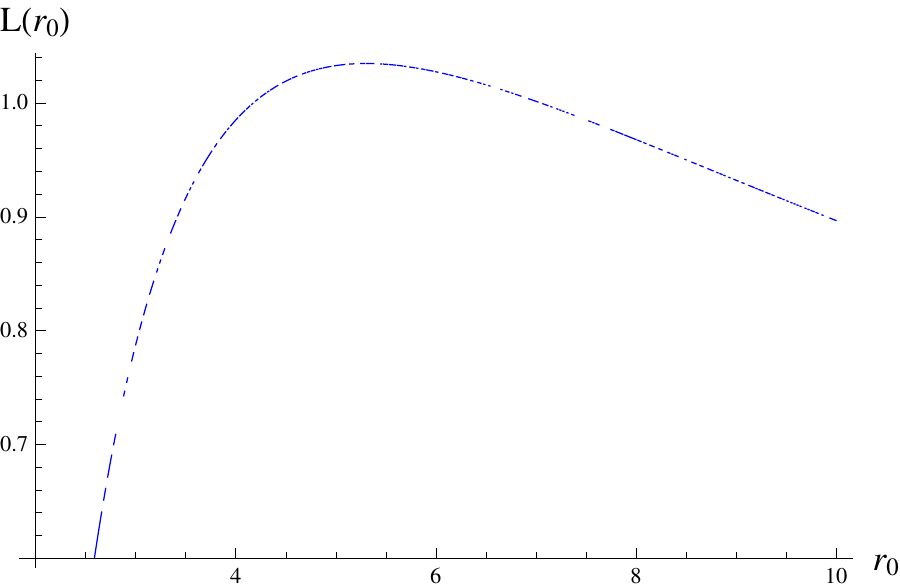}
  \captionof{figure}{Plot of $L(r_0)$ vs. $r_0$ for KT, $r_s=2$. }
  \label{fig:KTL3}
\end{minipage}%
\begin{minipage}{.5\textwidth}
  \centering
  \includegraphics[width=.9\linewidth]{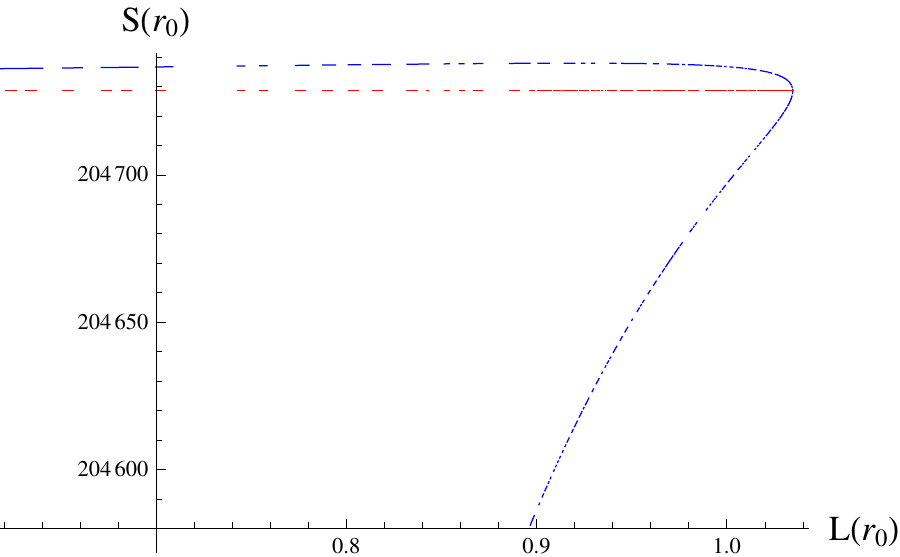}
  \captionof{figure}{Plot of $S(r_0)$ vs. $L(r_0)$ for KT, $r_s=2$. }
  \label{fig:KTSS3}
\end{minipage}
\end{figure}

As one can see the form of the plots are very similar to the KS geometry if the naked singularity is placed at small $r$ as in here it is set to $r_s=0.5$. However increasing $r_s$ causes that the red dashed line goes down and the butterfly shape of the diagram get a flatter curvature as it is shown in Figures \ref{fig:KTL2}, \ref{fig:KTSS2}, \ref{fig:KTL3} and \ref{fig:KTSS3}.

\subsection{Klebanov-Strassler}

For the KS background, the functions are
\begin{gather}
\alpha=h^{-\frac{1}{2}}(\tau), \ \ \ \ \ \beta(\tau)=\frac{h(\tau) \epsilon^{4/3}}{6 K^2(\tau)},\nonumber\\
H(\tau)=\frac{8\pi^6}{3} \epsilon^{20/6} h(\tau) K^2(\tau) \sinh^4(\tau).
\end{gather}
Again, $\beta$ is a monotonically decreasing function and $H$ is a monotonically increasing function. 
Now using $\beta$ and $H$ we study the entanglement entropy of a strip in this geometry.
So
\begin{gather}
L(\tau_0)=\frac{2^{\frac{5}{6}} \epsilon^{\frac{2}{3}} }{\sqrt{3}} \int_{\tau_0}^\infty d\tau \frac{\sinh(\tau)}{(\sinh(2\tau) -2\tau)^{\frac{1}{3}}}\sqrt{ \frac{h(\tau)} { \frac{\sinh(\tau)^2 }{\sinh(\tau_0)^2} \left( \frac{\sinh(2\tau) -2\tau}{\sinh(2\tau_0) -2\tau_0}  \right)^{\frac{2}{3} } -1 } },\nonumber\\
S_C(\tau_0)=\frac{V_2 \pi^3 \epsilon^4}{3G_N^{(10)}} \int_{\tau_0}^\infty d\tau \frac{h(\tau) \sinh^2(\tau)}{ \sqrt{1- \left( \frac{\sinh(\tau_0) }{\sinh(\tau) }  \right)^2 \left( \frac{\sinh(2\tau_0) -2\tau_0 }{\sinh(2\tau)-2\tau }  \right)^{\frac{2}{3}}  } },\nonumber\\
S_D(\tau_0)=\frac{V_2 \pi^3 \epsilon^4}{3 G_N^{(10)}} \int_{\tau_\Lambda}^\infty d\tau h(\tau) \sinh^4 \tau.
\end{gather}

\begin{figure}[h!]
\centering
\begin{minipage}{.5\textwidth}
  \centering
  \includegraphics[width=.9\linewidth]{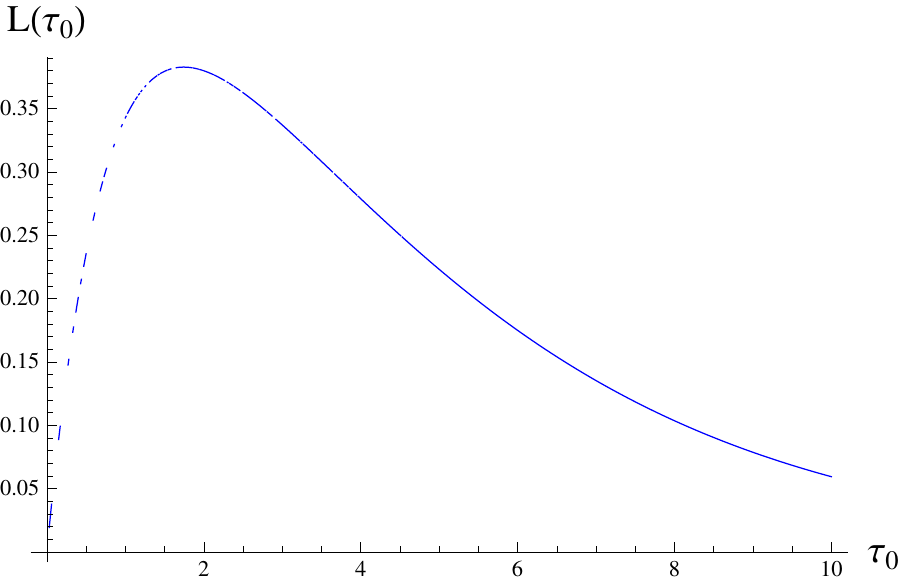}
  \captionof{figure}{Plot of $L(u_0)$ vs. $u$ for KS. }
  \label{fig:KSL2}
\end{minipage}%
\begin{minipage}{.5\textwidth}
  \centering
  \includegraphics[width=.9\linewidth]{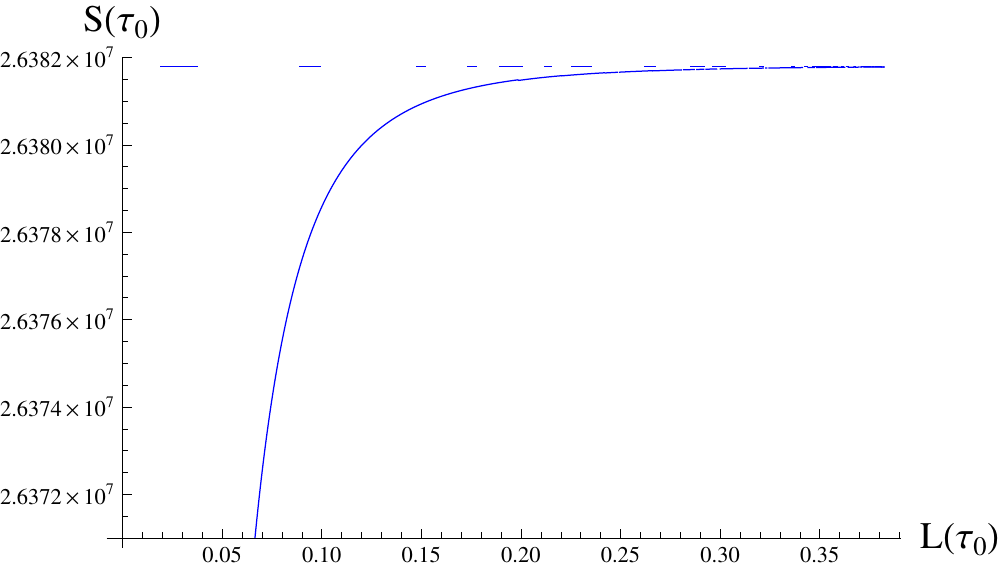}
  \captionof{figure}{Plot of $S(u_0)$ vs.$u_0$ for KS. }
  \label{fig:KSE2}
\end{minipage}
\end{figure}

The plot of the length of the connected solution, $L$ and the entanglement entropy, $S$ is shown in Figures \ref{fig:KSL2} and \ref{fig:KSE2}. The entanglement entropy of KS case with dynamical flavors were also studied in \cite{Georgiou:2015pia}. From the behavior of $L$ and the butterfly shape of $S$ one can detect the confining phase. One can see however that the phase transition is milder relative to the Witten QCD and KT cases. This milder phase transition were also seen for the Schwinger effect phase transition.

\subsection{Maldacena-Nunez background}
For the Maldacena-Nunez background, the functions are
\begin{gather}
\alpha(r)=e^{\phi}, \ \ \ \ \ \ \beta(r)=1, \ \ \ \ \ \ \ V_{\text{int}}=8\pi^3 e^{2h} e^{\frac{5}{2} \phi},\ \ \ \ \         H(r)=4\pi^6 e^{8\phi_0} (\sinh 2r)^4.
\end{gather}

\begin{figure}[h!]
\centering
\begin{minipage}{.55\textwidth}
  \centering
  \includegraphics[width=.9\linewidth]{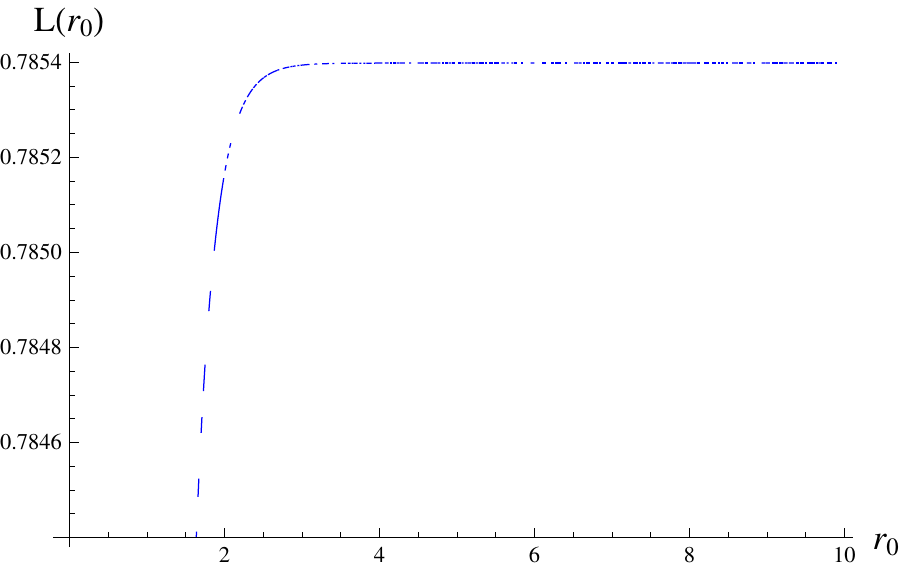}
  \captionof{figure}{Plot of $L(r_0)$ v.s $r_0$ for MN model. }
  \label{fig:LMN}
\end{minipage}%
\begin{minipage}{.55\textwidth}
  \centering
  \includegraphics[width=.9\linewidth]{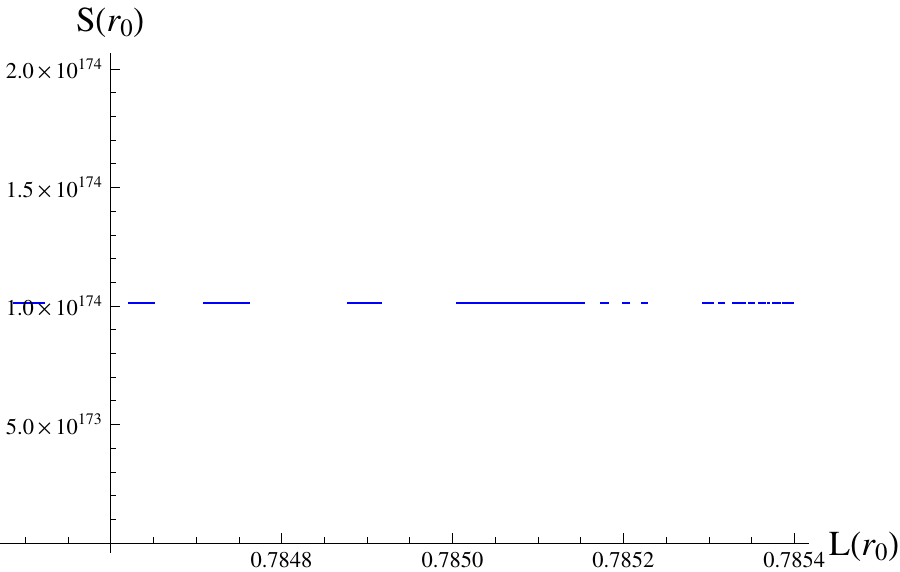}
  \captionof{figure}{Plot of $S$ vs. $L(r_0)$ for MN model. }
  \label{fig:SMN}
\end{minipage}
\end{figure}

Unlike the other backgrounds that we study here, for the Maldacena-Nunez metric, the function $\beta$ is a constant and is not monotonically decreasing. Also $H(r)$ is not monotonically increasing.
The $L(r_0)$ and EE functionals are

\begin{gather}
L(r_0)=\int_{r_0}^\infty dr \frac{2}{\sqrt{\frac{(\sinh 2r)^4}{(\sinh 2r_0)^4}-1 } },\nonumber\\
S_C(r_0)=\frac{V_2 \pi^3 e^{4\phi_0} }{G_N^{(10)}} \int_{r_0}^\infty dr \frac{(\sinh 2r)^2}{\sqrt{1-\frac{(\sinh 2r_0)^4}{(\sinh 2r)^4}} },\nonumber\\
S_D(r_0)=\frac{V_2 \pi^3 e^{4 \phi_0} }{G_N^{(10)}} \int_{r_0}^\infty dr (\sinh 2r)^2. 
\end{gather}
Their plots are shown in Figures \ref{fig:LMN} and \ref{fig:SMN}.

\subsection{Klebanov-Witten}
As the Klebanov-Witten geometry is not confining, it would be interesting to compare the behavior of the functions $\beta$ and $H$ and also $L$ and $S$ with the confining geometries studied above. 

\begin{figure}[h!]
\centering
\begin{minipage}{.5\textwidth}
  \centering
  \includegraphics[width=.9\linewidth]{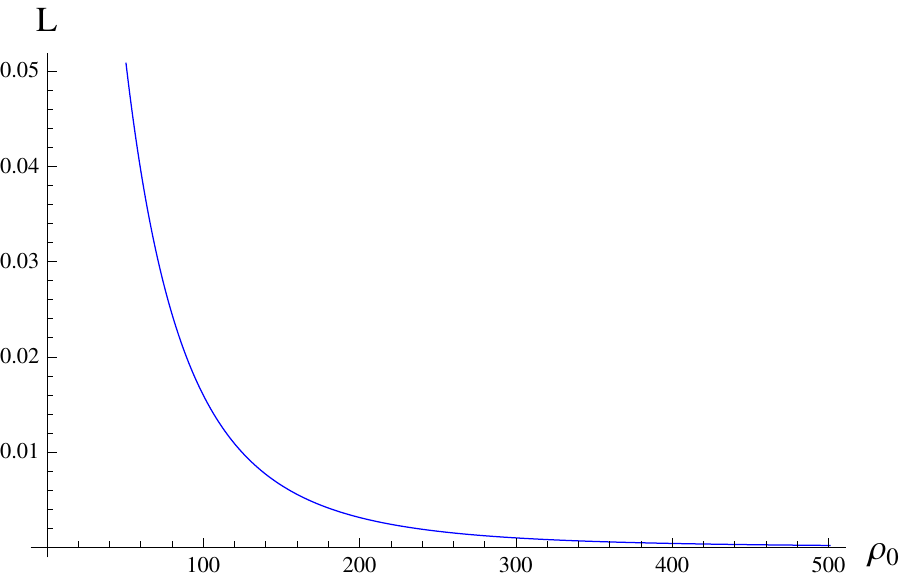}
  \captionof{figure}{Plot of $L$ vs. $\rho$ for KW. }
  \label{fig:LKW}
\end{minipage}%
\begin{minipage}{.5\textwidth}
  \centering
  \includegraphics[width=.9\linewidth]{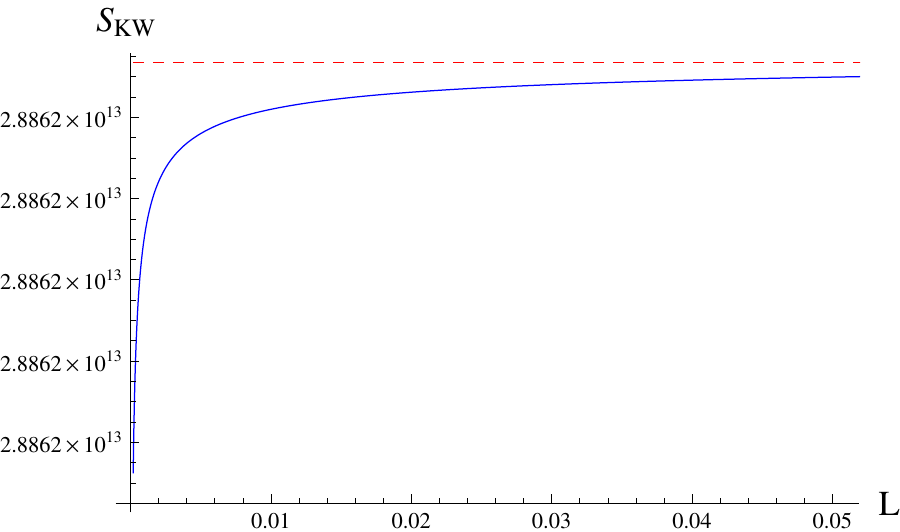}
  \captionof{figure}{Plot of $S$ vs. $L$ for KW. }
  \label{fig:KWEE}
\end{minipage}
\end{figure}

As one can see in figures \ref{fig:LKW} and \ref{fig:KWEE}, there is no phase transition in the plot of $S$ and the true solution is the connected one. Also there is no peak in the plot of $L$ which again specifies that KW is indeed a conformal geometry. This can be compared with the diagram of \ref{fig:11a}.

\section{The critical electric field in the presence of magnetic field}
Now in this section we assume that in addition to the electric field, a parallel and a perpendicular magnetic field components are also present. By using the Euler-Heisenberg Lagrangian, we then study the critical electric field which would lead to the Schwinger pair creation in our four confining geometries. We see that similar to the Sakai-Sugimoto and deformed Sakai-Sugimoto models, the parallel component would increase the pair creation rate and the perpendicular component would decrease it.

 \subsection{Maldacena-Nunez}
For the MN metric, we have
\begin{gather}
\sigma _{\text{string}-\text{MN}}=\frac{e^{\phi _0}}{2 \pi  \alpha ' \sqrt{2}}\frac{\sqrt{ \text{sinh}\left(2r_*\right)}}{\left(r^* \text{coth}\left(2r_*\right)-\frac{r_*{}^2}{\sinh^2 \left(2r_*\right)}-\frac{1}{4}\right){}^{\frac{1}{4}}}.
\end{gather}
Based on the assumption that a geometry has an IR wall and the probe D-branes would hit the wall, by calculating the imaginary part of the DBI action with a constant field strength, the authors in \cite{Hashimoto:2014yya} showed that the critical electric field, where the Euler-Heisenberg Lagrangian becomes imaginary and therefore the Schwinger effect start to take place, is at $E_c$ which is,
\begin{gather} \label{critical E}
E_{c}=\sigma_{\text{string}} \sqrt {\frac{{\sigma^2}_{\text{string}}+|\vec{B}|^2 }   {\sigma_{\text{string} }^2+|\vec{B}_{||}|^2 }  }.
\end{gather}
Here $\sigma_{\text{string}}=\frac {g (r_{*} ) } {2 \pi \alpha}$ and $r_{*}$ is where the DBI action vanishes.
By calculating the equation \ref{critical E},  one can get
\begin{gather}
E_{c-MN}=\frac{e^{\phi _0} \sqrt{\sinh  \left(2 r_*\right)}}{2 \pi  \alpha ' \left(4 \coth \left(2 r_*\right)r_* -1-4 \text{csch}^2\left(2 r_*\right) r_*^2\right){}^{\frac{1}{4}} } \times \nonumber\\  \left(\frac{e^{2 \phi _0} \sinh \left(2 r_*\right)+4\pi ^2 \alpha '^2B^2\text{  } \sqrt{4 \coth \left(2 r_*\right) r_*-1-4 \text{csch}^2\left(2 r_*\right) r_*^2} }{e^{2 \phi _0} \sinh \left(2 r_*\right)+4\pi ^2 \alpha '^2 {B_{||}}^2 \sqrt{4 \coth \left(2 r_*\right) r_*-1-4 \text{csch}^2\left(2 r_*\right) r_*^2} }\right)^{\frac{1}{2}}.
\end{gather}
For $\vec{B_{||}}=\vec{B}=0$ the critical $E$ is
\begin{gather}
E_{cr-MN}=\frac{e^{\phi _0} \sqrt{\sinh  \left(2 r_*\right)}}{2 \pi  \alpha ' \left(4 \coth \left(2 r_*\right)r_* -1-4 \text{csch}^2\left(2 r_*\right) r_*^2\right){}^{\frac{1}{4}} }.
\end{gather}

We can re-derive this result by finding the imaginary part of the Euler-Heisenberg action and therefore we can find an indicator for the universality of the equation \ref{critical E} for the confining geometries.

We then calculate the DBI action by finding the induced metric of MN on D3-brane or D7-brane profile. The D3-brane profile is
\begin{gather}
ds^2=(1+\frac{R^4}{r^4})^{-\frac{1}{2}}(-dt^2+d {\vec {x}}^2)+(1+\frac{R^4}{r^4})^{\frac{1}{2}} (dr^2+r^2 d{\Omega_5}^2), 
\end{gather}
and the D7-brane profile is
\begin{gather}
ds_{10}^2=\frac{r^2}{R^2} dx_\mu dx^\mu +\frac{R^2}{r^2} ds_{(6)}^2,\nonumber\\
ds_{(6)}^2=dr^2+\frac{r^2}{3} \big(\frac{1}{4} ({w_1}^2 +{w_2}^2 ) +\frac{1}{3}{w_3}^2+ (d\theta-\frac{1}{2} f_2)^2 +(\sin \theta d\phi -\frac{1}{2} f_1)^2 \big),
\end{gather}
where $R^4=\frac{27}{4} \pi g_s N_c {\ell_s}^4$ and $R$ is the $\text{AdS}_5$ radius.

Now we calculate the DBI action by finding the induced metric on D7-brane. So
\begin{gather}
\mathcal{L}_{\text{MN}-\text{D7}}=-T_7\int  d^2x\text{  }\text{d$\Omega $}_5 E^{-\phi _0}\int _{r_{\text{kk}}}^{\infty }dr \frac{\left(r \coth (2r)-\frac{r^2}{\sinh (2r)}-\frac{1}{4}\right)^{\frac{1}{4}}}{\sqrt{\frac{1}{2} \sinh (2r)}}\times \nonumber\\ \sqrt{\frac{2 e^{4 \phi _0} \sinh ^4(2 r)}{1-8 r^2-\cosh (4 r)+4 r \sinh (4 r)}-\frac{(2 \pi  \alpha )^2 e^{2 \phi _0} \left(F_{01}{}^2-F_{12}{}^2-F_{13}{}^2-F_{23}{}^2\right) \sinh (2 r)}{\sqrt{4 r \coth (2 r)-4 r^2 \text{csch}^2(2 r)-1}}-(2 \pi  \alpha )^4\text{  }F_{01}{}^2 F_{23}{}^2}.
\end{gather}
For $F_{01}=E_1, F_{23}=B_1, F_{13}=B_2, F_{12}=B_3$ we derive the Lagrangian as
\begin{gather}
\mathcal{L}_{\text{MN}-\text{D7}}=-T_7\int  d^2x\text{  }\text{d$\Omega $}_5 E^{-\phi _0}\int _{r_{\text{kk}}}^{\infty }dr \frac{\left(r \coth (2r)-\frac{r^2}{\sinh (2r)}-\frac{1}{4}\right)^{\frac{1}{4}}}{\sqrt{\frac{1}{2} \sinh (2r)}}\times  \nonumber\\ \sqrt{\frac{2 e^{4 \text{$\phi $0}} \sinh ^4(2 r)}{1-8 r^2-\cosh (4 r)+4 r \sinh (4 r)}+\frac{(2 \pi  \alpha )^2e^{2 \text{$\phi $0}} \left({\vec{B}}^2-E_1{}^2\right)\text{  }\sinh (2 r)}{\sqrt{4 r \coth (2 r)-4 r^2 \text{csch}^2(2 r)-1}}-(2 \pi  \alpha )^4 B_1{}^2 E_1{}^2}.
\end{gather}
By solving the $E$ which makes this Lagrangian imaginary, we find the same $E$ as equation \ref{critical E}. So this way the universality of this relation can be checked again.
\begin{figure}[h!]
  \centering
    \includegraphics[width=0.5\textwidth]{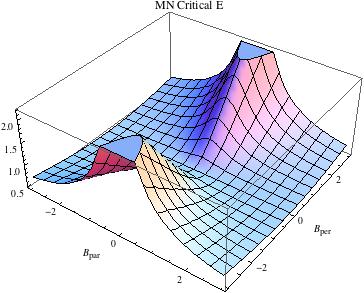}
       \caption{The $E_{cr}$ versus parallel and perpendicular magnetic fields for $\alpha=1$, $\phi_0=0$ and $r=2$.}
       \label{fig:critical}
\end{figure}
From Figure \ref{fig:critical} one can see that the behavior of the critical $E$ vurses the magnetic fields is very similar to the general figure as in \cite{Hashimoto:2014yya}.  
\begin{figure}[h!]
  \centering
    \includegraphics[width=0.5\textwidth]{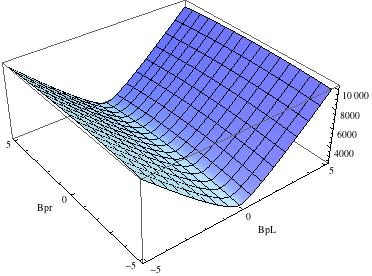}
       \caption{The $Im \mathcal{L}$ for the MN background, vs. parallel and perpendicular magnetic fields for $\alpha=1$, $\phi_0=0$ and $E=10$. (We have normalized the fields.) }
\end{figure}

One can specifically check that for this background, by increasing the parallel magnetic field the imaginary part of the EH Lagrangian would increase which leads to an increase in the pair creation, but increasing the perpendicular magnetic field decreases the rate. 

Now we look at the electric field dependence. For the case of $B_{||}=0$, we have
\begin{gather}
Im \mathcal{L} _{perp.B} = \int_{0.05}^5  dr \left(r \coth (2 r)-r^2 \text{csch}(2 r)-\frac{1}{4}\right)^{\frac{1}{4}} \times \nonumber\\  \sqrt{\frac{4 \sinh ^3(2 r)}{8 r^2+\cosh (4 r)-4 r \sinh (4 r)-1}-\frac{2 (2\pi  \alpha )^2 \left(B_{\text{pr}}{}^2-E_{\|}{}^2\right)\text{  }}{\sqrt{4 r \coth (2 r)-4 r^2 \text{csch}^2(2 r)-1}}},
\end{gather}
\begin{figure}[h!]
  \centering
    \includegraphics[width=0.5\textwidth]{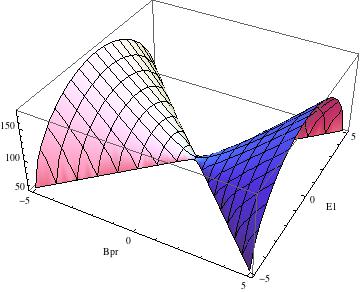}
       \caption{The $Im \mathcal{L}_{MN}$ versus electric field and perpendicular magnetic fields for  $B_{||}=0$, $\alpha=1$, $\phi_0=0$ and $r=2$  .}
\end{figure}
and for the case of $B_{\perp}=0$, we have
\begin{gather}
Im \mathcal{L} _{para.B} = \int_{0.05}^5  dr \left(r \coth (2 r)-r^2 \text{csch}(2 r)-\frac{1}{4}\right)^{\frac{1}{4}}\times \nonumber\\  \sqrt{\frac{2(2\pi \alpha )^4 B_{\|}{}^2 E_{\|}{}^2}{\sinh (2 r)}+\frac{4 e^{4 \phi _0} \sinh ^3(2 r)}{8 r^2+\cosh (4 r)-4 r \sinh (4 r)-1}-\frac{2(2\pi \alpha)^2 e^{2 \phi _0} \left(B_{\|}{}^2-E_{\|}{}^2\right)}{\sqrt{4 r \coth (2 r)-4 r^2 \text{csch}^2(2 r)-1}}}.
\end{gather}

\begin{figure}[h!]
  \centering
    \includegraphics[width=0.5\textwidth]{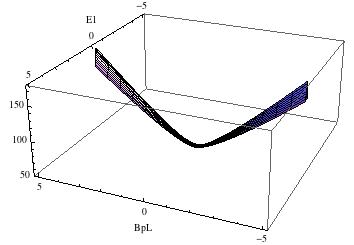}
       \caption{The $Im \mathcal{L}_{MN}$ versus electric field and parallel magnetic fields for $B_{\perp}=0$,  $\alpha=1$, $\phi_0=0$ and $r=2$.}
\end{figure}

\subsection{D3 probe brane in MN background}
The pullback of the Maldacena Nunez metric (which consists of N D5 brane)  on D3-brane worldvolume is $e^\phi \eta_{\mu\nu}$. So the Lagrangian would be
\begin{gather}
\mathcal{L}_{\text{MN}-\text{D3}}=-T_3\int  d^4x\text{  }\surd (e^{-4 \phi _0} (-(2\pi  \alpha )^4 e^{4 \phi _0} F_{01}{}^2 F_{23}{}^2+(-1+4 r \coth (2 r)) \text{csch}^2(2 r)-4 r^2 \text{csch}^4(2 r)- \nonumber \\ (2\pi  \alpha )^2 e^{2 \text{$\phi $0}} (F_{01}{}^2-F_{12}{}^2-F_{13}{}^2-F_{23}{}^2) \text{csch}(2 r) \sqrt{-1+4 r \coth (2 r)-4 r^2 \text{csch}^2(2 r)})).
\end{gather}
There is no integral of ``r" for the D3-brane case, so we can actually study the Schwinger effect in different $r$ by inserting D3-brane in the geometry as the probe. 

\begin{figure}[h!]
  \centering
    \includegraphics[width=0.5\textwidth]{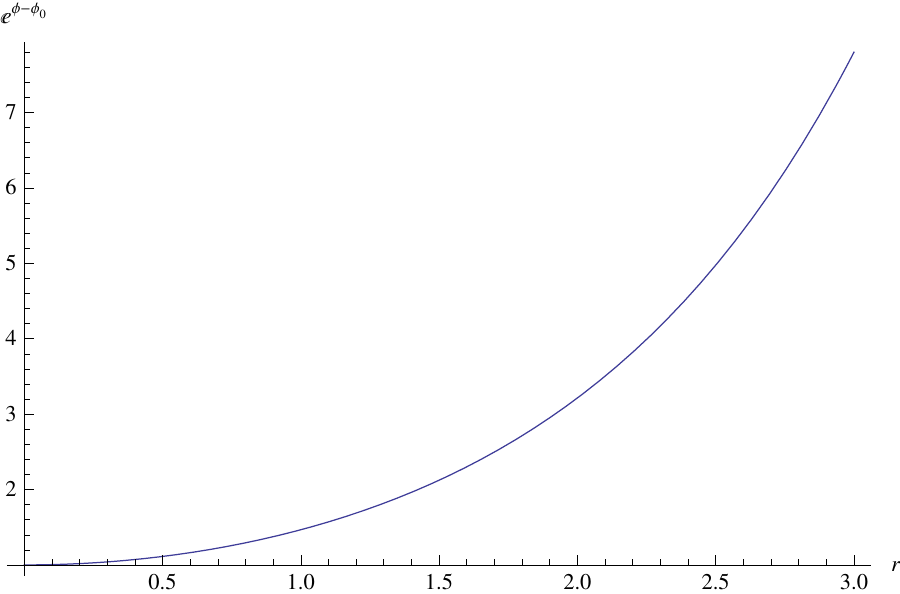}
       \caption{The potential on the D3 brane when all the fields are off.}
       \label{fig:curve}
\end{figure}

First we study the potential in MN geometry when all the electric and magnetic fields are off which is shown in Figure \ref{fig:curve}.

As one can see, all the D3 branes are being pulled to the tip of $r = 0$ when they are inserted in MN and KS, due to the effect of $F_5$ and gravity. 

Now we investigate the potential when the fields are on.
If we take $F_{01}=E_1, F_{12}=B_3, F_{23}=B_1 , F_{13}=B_2$ then the potential is
\begin{gather}
V= \Big (e^{-4 \phi _0} (-(2\pi  \alpha )^4 B_1{}^2 e^{4 \phi _0} E_1{}^2+\text{csch}(2 r) ((2\pi  \alpha )^2 e^{2 \phi _0} (B_1{}^2+B_2{}^2+B_3{}^2-E_1{}^2) \times \nonumber\\ \sqrt{-1+4 r (\coth (2 r)-r \text{csch}^2(2 r))}+\text{csch}(2 r) (-1+4 r (\coth (2 r)-r \text{csch}^2(2 r))))) \Big ){}^{\frac{1}{2}}.
\end{gather}

\begin{figure}[h!] 
  \centering
    \includegraphics[width=0.5\textwidth]{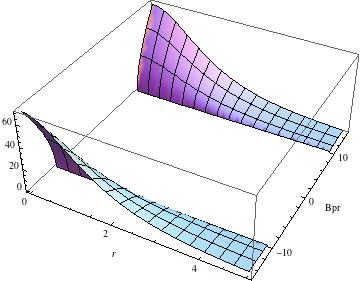}
       \caption{The potential on the D3 brane for $E_{||}=10, B_{||}=0, \alpha=1, \phi_0=0$. }
       \label{fig:pot}
\end{figure}
This is shown in Figure \ref{fig:pot}. One can see that the potential is looking like two walls which have higher slopes near the IR region ($ r\to 0$). In the UV ($r\to \infty$) the potential is zero and has zero slope. Increasing $E_{||}$ would increase the slope of the potential and the pair creation rate.

\begin{figure}[h!]
  \centering
    \includegraphics[width=0.5\textwidth]{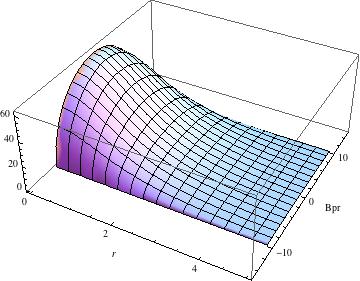}
       \caption{The imaginary part of Lagrangian related to the pair creation for $E_{||}=10, B_{||}=0,  \alpha=1, \phi_0=0$. }
        \label{fig:Impot}
\end{figure}

Again from Figure \ref{fig:Impot}, one can notice that the pair creation happens with higher rate near the origin, at $r \to 0$, and at bigger $r$ it would decrease. Increasing the perpendicular magnetic field would decrease the pair creation till make it zero at a specific value of the perpendicular magnetic filed at any $r$.

Now we turn off the perpendicular magnetic field and study the effect of the parallel component of the magnetic field. As it can be seen from the following three figures, parallel magnetic field generally increases the whole pair creation rate, however at $B_{||}=0$ the pair creation is mainly happening at small $r$. Increasing $B_{||}$ increases the pair creation in the UV and decreases the area which the pair creation is happening in the IR till makes it zero there, but in total, increasing $B_{||}$ would increase the imaginary part of Lagrangian and therefore the rate of the Schwinger effect.

\begin{figure}[h!]
  \centering
    \includegraphics[width=0.5\textwidth]{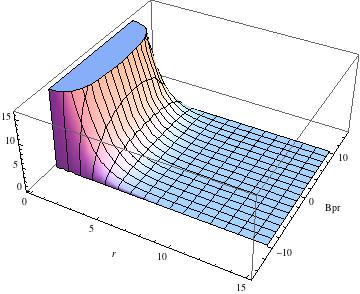}
       \caption{The imaginary part of the Lagrangian for $E_{||}=10, B_{||}=0, \alpha=1, \phi_0=0$. }
       \end{figure}
       \begin{figure}[h!]
  \centering
    \includegraphics[width=0.5\textwidth]{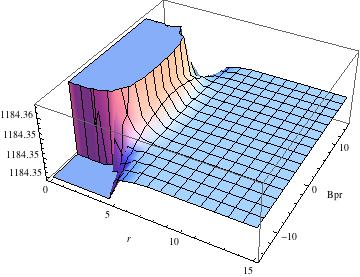}
       \caption{The imaginary part of Lagrangian for $E_{||}=10, B_{||}=3, \alpha=1, \phi_0=0$. }
       \label{fig:interB}
\end{figure}

It would be interesting to notice that when $E_{||}=0$ even with a strong $B_{||}$, no pair creation would happen. Only at small $r$, in the IR region, increasing the perpendicular magnetic field would increase the potential. 

One should notice that heavier states with higher charges lie in the IR as the hadrons' wave functions fall as $r^{-\Delta}$ and $\Delta \sim J$. This can be one reason that we see such a behavior in the IR. One should also note that Wilson loops also find it more favorable to lie at the end of the space in the IR. 

\begin{figure}[h!]
  \centering
    \includegraphics[width=0.5\textwidth]{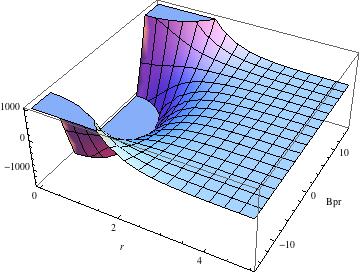}
       \caption{The potential for $E_{||}=10$ and $B_{||}=0$ showing the hole in the IR region.}
       \label{fig:Potz}
\end{figure}

The plot of $- Det(F_{MN} + g_{MN})$  vs. $r$ and $B_{\perp}$, for a constant $E = 10$, and zero parallel magnetic field, Figure \ref{fig:Potz} , shows that there exist a critical $r$, ( for $B_{\perp} =0$ is around $r= 1.2$ ) where a hole is forming.  For confining theories there is a critical $r_0$ where $\frac{\partial g_{tt} }{ \partial r} \big |_{r=r_0}=0$ which for our specific geometry gives $r=1.118$ which is close to what we have seen from the figures. Increasing $E$ would increase the radius of this hole.

\subsection{Klebanov-Strassler}
The Klebanov-Strassler (KS) metric is 
\begin{gather}
ds_{10}^2=h^{-\frac{1}{2}}(\tau) dx_\mu dx^\mu+h^{\frac{1}{2}}(\tau) {ds_6}^2.
\end{gather}
The form of this metric is also like \ref{eq:form}, so the D-branes hit the IR wall. The Klebanov-Strassler string tension is
\begin{gather}
\sigma _{\text{string}-\text{KS}}=\frac{h^{\frac{-1}{2}}\left(\tau _*\right)}{2 \pi  \alpha '}.
\end{gather}
Using this and the equation \ref{critical E}, we can find the critical $E$ as 
\begin{gather}
E_{cr}=\frac{1}{2 \pi  \alpha '}\sqrt{\frac{\frac{1}{h(\tau )}+\left( 2\pi  \alpha ' B\right)^2 }{1+\left(2\pi  \alpha 'B_{\|}\right){}^2 h(\tau )}}.
\end{gather}
For $\vec{B}=\vec{B_{||} }=0$, as it was obvious from $E_{cr}=\frac{g(r_*)}{2 \pi \alpha^\prime} $, we would have  $E_{cr}=\frac{h^{-\frac{1}{2} }(\tau)}{2 \pi \alpha^\prime}$.

\begin{figure}[h!] \label{fig:h}
  \centering
    \includegraphics[width=0.5\textwidth]{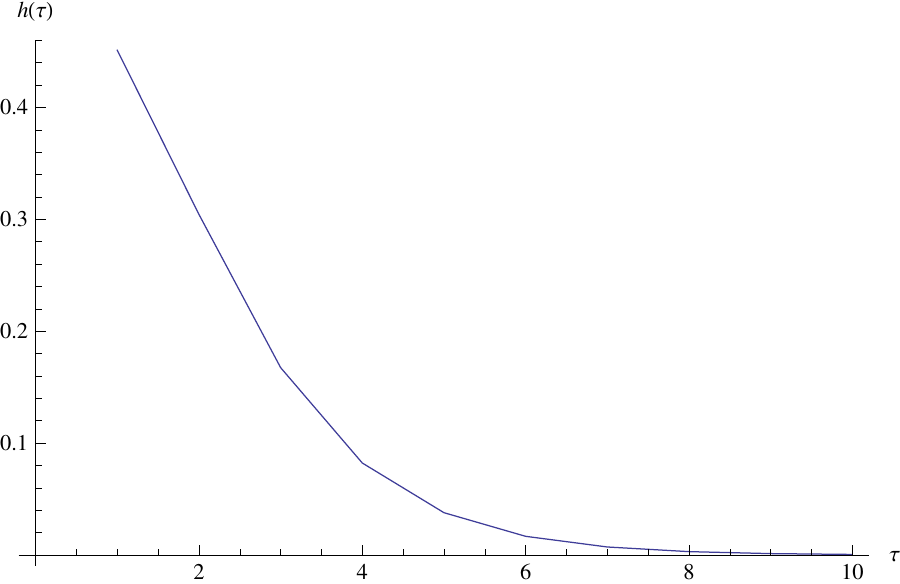}
       \caption{The behavior of the function $h(\tau)$ vs. $\tau$ for the Klebanov-Stressler metric. }
       \label{fig:hh}
\end{figure}

The behavior of the function $h(\tau)$ is shown in Figure \ref{fig:hh}. It shows that after $\tau >10$ it is practically zero. So for showing the numerical plot we don't actually need to take the integral of the Lagrangian to $\tau=\infty$, but rather we take as $\tau=10$.
So the Lagrangian for the constant field strength is 
\begin{gather}
\mathcal{L}_{\text{KS}}=-T_7\int  d^4x\text{  }\text{d$\Omega $}_3 e^{-\phi _0}\int _{\tau _{\text{kk}}}^{\infty }d\tau \sqrt{\frac{1+(2\pi \alpha)^2 \left(B_{||}{}^2+B_{\perp}{}^2-E_{\|}{}^2\right)  h(\tau )-(2\pi \alpha)^4 B_{\text{pL}}{}^2 {E_{||}}^2  h(\tau )^2}{h(\tau )^2}}.
\end{gather}

\begin{figure}[h!] \label{fig:h}
  \centering
    \includegraphics[width=0.5\textwidth]{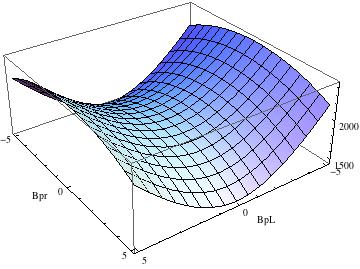}
       \caption{ The Im $\mathcal{L}_{KS}$ vs. $B_{\perp}$ and $B_{||}$ of the KS background for $\tau=9.5$, $h(\tau)= 0.000798174$, and $E_{||}=10$, $\alpha=1$. }
       \label{fig:IMKS}
\end{figure}

Again as in Figure \ref{fig:IMKS} one can check that for the KS background by increasing the parallel magnetic field the imaginary part of the Lagrangian would increase leading to an increase in pair creation and vice versa for the perpendicular magnetic field, so by increasing the perpendicular magnetic field, the imaginary part of the Lagrangian would decrease leading to a more stable phase with lower rate of pair creation.

\begin{figure}[h!] \label{fig:h}
  \centering
    \includegraphics[width=0.5\textwidth]{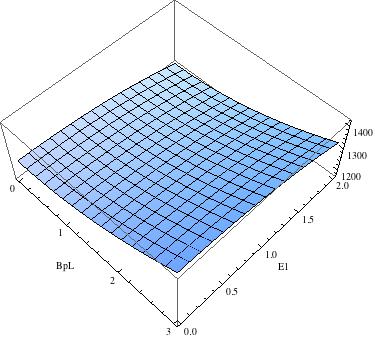}
       \caption{The Im $\mathcal{L}_{KS}$ vs. $B_{\perp}$ and $E$ of the KS background for $\alpha=1$, $\tau=9.5$, $h(\tau)= 0.000798174$. }
\end{figure}

\subsection{Witten-QCD}

The Witten-QCD metric is
\begin{gather}
ds^2=\left(\frac{u}{R}\right)^{\frac{3}{2}}(\eta_{\mu\nu} dx^\mu dx^\nu)+\left(\frac{R}{u}\right)^{\frac{3}{2}}\frac{du^2}{f(u)}+\left(\frac{u}{R}\right)^{\frac{3}{2}}\frac{4R^3}{9u_0} f(u) d\theta^2+R^{\frac{3}{2}} u^{\frac{1}{2}} d\Omega_4^4.
\end{gather}
Since the generic form of the confining geometries is 
\begin{gather} \label{eq:form}
ds^2=g(r) \eta_{\mu\nu} dx^\mu dx^\nu+f(r) dr^2 +h(r) [\text{internal space}],
\end{gather}
in the Witten QCD model, the internal geometry mix with the radial coordinate $u$, but still it is confining and the probe D-branes hit the IR wall. 

For WQCD, $\sigma_{st}=\frac{1}{2\pi \alpha'} (\frac{u_*}{R})^{\frac{3}{2}}$ which is consistent with our calculation of the potential in the previous sections.

Assuming $T_F=\frac{1}{2\pi \alpha'}$ from the above equation one would get
\begin{gather}
E_c=T_F\left (\frac{u_{*}}{R}\right)^{\frac{3}{2}} \sqrt{\frac{T_F^2 (\frac{u_*}{R} )^3+\vec{B}^2  }{T_F^2 (\frac{u_*}{R} )^3+\vec{B_{||}}^2} },
\end{gather} 
and when the magnetic field is off, this leads to the familiar result $E_c=\sigma_{st}$. Now the DBI action in D5-brane background including a constant electromagnetic field is
\begin{gather}
S_{D5}^{\text{DBI}} =-T_5 \int d^4 x du e^{-\phi} \sqrt{-\text{det}( P[g]_{ab}+2\pi \alpha' F_{ab})}=\nonumber\\
-\frac{2}{3} \frac{T_5 g_s}{R^{\frac{3}{4}}} \int_{u_{KK}} ^\infty du \sqrt{ \frac{u^6+4(B_\perp^2+B_\parallel^2-E^2)\pi^2 R^3 u^3 \alpha^2-16 B_\parallel^2 E^2 \pi^4 R^6 \alpha^4 }{f R^3 u_0} }.
\end{gather}

Again one can see that in this model too, by increasing the parallel magnetic field the imaginary part of the Lagrangian and therefore the rate of pair creation would increase, while increasing the perpendicular magnetic field would decrease it.

\subsection{Klebanov-Tseytlin}
The critical electric field in the presence of a magnetic field in KT background would be
\begin{gather}
E_c=T_F \frac{L^2}{r_*^2} \sqrt{\ln \frac{r_*}{r_s} } \sqrt{ \frac{T_F^2 \frac{L^4}{r_*^4 } \ln \frac{r_*}{r_s} + |\vec{B}|^2  }{T_F^2 \frac{L^4}{r_*^4} \ln \frac{r_*}{r_s} +| \vec{B}|^2}  },
\end{gather}
and the DBI action in D5-brane background is 
\begin{gather}
S_{D5}^{DBI}= -\frac{T_5}{L^2} \int_{r_{KK}}^\infty dr  \sqrt{ \frac{r^8 +4(B_\perp^2+B_\parallel^2-E^2) L^4 \pi^2 r^4 \alpha^2 \ln \frac{r}{r_s} -16 B_\parallel^2 E^2 L^8 \pi^4 \alpha^4 (\ln(\frac{r}{r_s})  )^2 }{r^2  \ln \frac{r}{r_s} } },
\end{gather}
which again one can check that increasing the parallel magnetic field would increase the rate of pair creation and increasing the perpendicular magnetic field decreases the rate.

\section{Discussion}
In this work we studied the Schwinger effect and the entanglement entropy in four supergravity confining geometries; the Witten-QCD, the Maldacena-Nunez, the Klebanov-Tseytlin and the Klebanov-Strassler model. First we derived the PDE equations outlining the string profile which is moving in the bulk of these models. Solving these PDE equations analytically, and then calculating the proper area between the event horizon on the profile and the boundary can result in the free energy of accelerating particles. If one also finds the Unruh temperature of accelerating particles in these confining supergravity backgrounds, then in the semi-classical limit one can find the entanglement entropy of pairs of particles moving on specific trajectories.

 We then studied the electric potentials in these models. By using the DBI action we derived the critical electric field, and then by using a probe D3-brane, we derived the total potential and the distance between the quark and anti quark in each background. By solving the integrals numerically, we plotted the phase diagram of the total potential versus the distance between the quarks and anti quarks, $x$, in each geometry. As was expected we found three similar phases in each background. For the small electric fields or $\alpha$ no pair creation would be present. By increasing the electric field the Schwinger pair creation would happen by a tunneling process and with an exponential suppression. Finally by increasing the electric field the potential in all of these models becomes unstable catastrophically where the probability is no longer exponentially suppressed. The rate of this phase transition is faster for WQCD and KT, relative to KS and MN. We also presented the phase diagram of the Klebanov-Witten and AdS for comparing the results with the conformal cases. In the non-confining geometries the first phase does not exist.

We then studied the phase diagram of the entanglement entropy of a strip in these confining geometries. We found the expected butterfly shape in the diagram of the entropy versus the length of the connected solution and a peak in the diagram of the length versus the minimal radial position in all of these backgrounds indicating the existence of a confinement phase. Again we found that the rate of changing phases is faster for WQCD and KT relative to KS and MN. It would be interesting to further study the possible relations between the Schwinger effect and the entanglement entropy. Finding such a relation could be very useful, as it can be used to measure the entanglement entropy in the lab with many applications in the condensed matter systems. 

In the final section we were interested to study the effects of an additional magnetic field on the rate of pair creation in each of these models. We found the critical electric field which makes the pair creation happening in the presence of the magnetic field and then by calculating the imaginary part of the Euler-Heisenberg effective Lagrangian, we studied the rate of pair creation for each model. The results was that similar to the Sakai-Sugimoto and deformed Sakai-Sugimoto models, \cite{Hashimoto:2014yya}, the parallel magnetic field to the electric field increases the pair creation rate and the perpendicular magnetic field decreases the rate. This could be a universal feature of all confining geometries which can be demonstrated analytically in the future works.

 \section*{Acknowledgements}
I would like to thank Leopoldo Pando Zayas for introducing me to this problem and Zahra Rezaei, Walter Tangarife, Elena Caceres and Dariush Kaviani for useful discussions and also the referee for her/his useful comments. I thank the Galileo Galilei Institute for Theoretical Physics and Institute for Research in Fundamental Sciences (IPM) for the hospitality and the INFN for the partial support during the completion of this work.

\providecommand{\href}[2]{#2}\begingroup\raggedright\endgroup

\end{document}